# A Computational Analysis and Visualization of In-Text Reference Networks Across Philosophical Texts

Robert Becker[1]*, Aron Culotta[1]

[1]Department of Computer Science, Tulane University

*Corresponding author: rbecker@tulane.edu

**Abstract**

We applied computational methods to analyze references across 2,245 philosophical texts, spanning from approximately 550 BCE to 1940 AD, in order to measure patterns in how philosophical ideas have spread over time. Using natural language processing and network analysis, we mapped over 294,970 references between authors, classifying each reference into subdisciplines of philosophy based on its surrounding context. We then constructed a graph, with authors as nodes and in-text references as edges, to empirically validate, visualize, and quantify intellectual lineages as they are understood within philosophical scholarship. For instance, we find that Plato and Aristotle alone account for nearly 10% of all references from authors in our dataset, suggesting that their influence—while widely recognized—may still be underestimated. As another example, we support the view that St. Thomas Aquinas served as a synthesizer between Aristotelian and Christian philosophy by analyzing the network structures of Aquinas, Aristotle, and Christian theologians. Our results are presented through an interactive visualization tool, allowing users to dynamically explore these networks, alongside a mathematical analysis of the network's structure. Our methodology demonstrates the value of applying network analysis with in-text references to study a large collection of historical works.

**Introduction**

Philosophy is as much a study of fundamental questions as it is a study of particular thinkers, and much of philosophical scholarship involves understanding the complex relationships between these thinkers over time. This work is primarily done through a deep qualitative analysis of texts, arguments, and historical contexts. A Kant scholar, for example, would have a rich understanding of how Plato, Aristotle, Leibniz, and Hume shaped Kant's thinking, as well as how his work influenced future philosophers. However, this understanding often becomes less certain with each degree of separation beyond one's expertise—the same scholar, for instance, may be less confident about the precise ways Hegel influenced Marx, or with how Schopenhauer influenced Nietzsche.

While traditional philosophical scholarship excels at interpreting and contextualizing these influences, it struggles to precisely measure them at scale. Mapping thousands of texts manually is impractical, and subjective interpretation can lead to inconsistencies in assessing influence. Though some philosophical connections—such as Aristotle's impact on medieval scholasticism (Baschera, 2009)—are well-documented, others remain debated, fragmented, or overlooked. Current tools to quantify influences across large bodies of text often rely on constructing networks through formal citations, which ignore other instances of in-text references to authors. This approach excels at investigating contemporary academic works which have a standardized citation format, but struggles to capture influence patterns across historical texts, where authors are more frequently referenced by in-text mentions of their name.

To address these gaps, our research presents a computational approach to deriving and analyzing reference networks across 2,245 philosophical texts—the largest computational analysis of the Western philosophical canon to date. We capture in-text mentions of authors to construct our reference network, enabling us to conduct network analysis on a historical corpus which lacks consistently formatted citations. We also classify each reference into subdisciplines of philosophy—such as ethics, metaphysics, or epistemology—by extracting the context in which each reference occurs. This adds a new dimension to our analysis of historical influence, allowing us to more precisely measure how influence has developed across different philosophical discussions.

In alignment with previous works, we view references as an effective and convenient proxy for intellectual influence—or at minimum, philosophical connection—when a sufficient amount of data is considered (Petrovich, 2024; Chi, 2022). The nature of this influence can take a variety of forms: an author could praise a previous thinker, as Nietzsche often does to Schopenhauer, or develop fierce criticisms of them, as Nietzsche does to Kant. An author could refine and expand upon previous arguments, as Leibniz does with Descartes' arguments for the existence of God, or an author could address the questions raised by a previous philosopher but offer different solutions, as Kant does with Hume. It is also common for canonical figures to synthesize ideas from competing schools, or for writers to adapt ideas into new areas, as Marx brought Hegel's dialectic method into his political analysis. As such, our findings are informed by philosophical literature to interpret the precise meaning of our reference metrics within its historical context.

Our paper proceeds in the following structure: In our next section, we review relevant literature within the digital humanities, network analysis, and academic philosophy. We then outline how our approach expands previous efforts in applying network analysis to historical texts on a large scale. In Section 3, we detail our pipeline for collecting data, removing noise, creating our network, classifying references, and developing our visualization tool. We also use this section to outline the models and algorithms selected for our analysis. In Section 4, we derive computational metrics from our reference network and consider their implications in characterizing the development of philosophy. Then, in Section 5, we investigate subsets of our network to explore the influence of specific philosophers as they are understood within philosophical literature. Finally, in Section 6, we share how our framework could be expanded to include contemporary philosophers, philosophers without surviving works, and non-Western lineages of philosophy, and share promising areas where Large Language Models could be applied to this area of research.

**Previous Work**

*2.1 Citation Analysis in Philosophy*

Previously, scholars have represented citations as graphs to examine trends and patterns across various fields. Most of this work exclusively uses formal citations, as they provide a reliable format to determine the source and authors of the referenced works. For example, Chi and Conix (2022) use these methods to measure the level of isolation between several texts, while Petrovich and Viola (2024) use a similar approach to understand the interconnection between neuroscience, philosophy, and psychology.

When graphs of informal citations are used, they are generally applied on a small number of works. For instance, many authors study conversation networks by representing authors as nodes and references as edges. These works are used to advance innovative literary interpretations, such as in Paradise Lost (Ruegg and Lee, 2020) and in Shakespeare's Hamlet (Kwon and Shim, 2017). They have also been used to study the role of figures in religious texts, such as in the Talmud (Satlow and Sperling, 2024) and the Old Testament (Lee and Webster, 2024). Other innovative representations of networks have also been used, such as Alfano's (2017), where concepts are represented as nodes and co-occurrences as edges.

*2.2 Topic Modelling Approaches in the Digital Humanities*

The use of computational methods to classify the topic of different texts is another common approach for analyzing large collections of historical texts. The traditional approach relies on Latent Dirichlet Allocation (Blei et al., 2003, Jelodar, 2019), where authors predefine a set of topics and assign words that indicate a piece of text should be classified under that topic. Other traditional approaches, such as Non-negative Matrix Factorization (Lee and Seung, 1999), fall under this general paradigm, with subtle differences in the underlying math.

Since 2014, scholars have begun exploring an additional paradigm for topic classification by relying on deep learning models to represent words as high dimensional embeddings (van Boven and Bloem, 2022; Wang et al., 2020; Herbelot et al., 2012). These approaches can better account for the context in which words occur while also enabling other methods for computational analysis, such as through visualizing text via dimensionality reduction (Devlin et al., 2019). As models for representing text as embeddings have become more sophisticated, the trade-offs between traditional approaches and deep learning approaches have become more pronounced: the former allows topic classification to remain simple and interpretable, while the latter offers potentially higher accuracy through less interpretable algorithms. More recently, researchers have experimented with using Large Language Models to systematically classify texts (Brown et al., 2024). This takes the dichotomy further, allowing scholars to apply topic modelling with greater precision at the cost of increased complexity in the underlying algorithm.

Irrespective of the technique used for topic modelling, there is a diverse array of applications within the digital humanities. Often, an entire text is classified, as Greene et al. (2024) do to analyze literary interviews in the Paris Review. Other approaches segment text into several portions which are classified separately, such as Shöch's work (2017) analyzing classical

French plays. Hierarchical clustering is used to classify text into both smaller and larger portions, which can reveal how the topics of smaller segments of text work to characterize the composition of those segments.

A common use case of topic modelling is to show how the discussion of subjects has increased and declined across time periods and geographic areas. When combined with the citation analysis methods discussed earlier, scholars have characterized and visualized the relationships between different fields and schools of thought. In philosophy, scholars have utilized these methods to analyze the rise and fall of certain topics in Ancient Greek writings (Köntges 2022), in contemporary philosophical journals (Malaterre and Lareau 2022), and in Hindu texts (Chandra and Rajan 2022).

*2.3 Traditional Philosophical Scholarship*

Much of the standard philosophy curriculum is both influenced by and reflected in Bertrand Russell's (1945) History of Western Philosophy, which outlines many canonical figures who are studied today. As is the nature of the philosophy discipline, the canon is constantly criticized and evaluated (Shapiro, 2016); some advocate for greater recognition of individual thinkers (Hopkins, 2011); others argue for the inclusion of more women, as well as non-white and non-Western thinkers. But while curricula have made gradual efforts to expand in recent decades, the core canon—from Plato to Kant—has remained the dominant framework in Western academic philosophy for centuries.

Within more specific threads, there is often debate over the precise ways in which thinkers have influenced others and how these influences should inform pedagogy. Leiter (2013), for example, writes about how Nietzsche's radical stylistic departure from philosophical traditions had a greater impact than is typically understood. The debate over the canon is also often directly engaged with by the canonical thinkers themselves. Nietzsche argued that Kant "clung to his university, submitted himself to its regulations, retained the appearance of religious belief, endured to live among colleagues and students: so it is natural that his example has produced above all university professors and professional philosophy" (Nietzsche, 1874). Russell (1945) acknowledged the influence of Aristotle's logic, but argued that dogmatic devotion to his system caused great harm: "Throughout modern times, practically every advance in science, in logic, or in philosophy has had to be made in the teeth of opposition from Aristotle's disciples." Hegel proposed that philosophy develops in a dialectic process, where a status quo (thesis) is challenged by a new idea (an antithesis), ultimately resulting in a new formulation of a novel concept (synthesis) (Forster, 1993).

*2.4 Philosophy's Engagement with the Digital Humanities*

Scholars with strong philosophical backgrounds often offer innovative approaches and interpretations when applying natural language processing techniques to texts within their domain expertise. Weatherson, for example, uses topic modelling to measure the proportion of philosophy articles published from 1876 to 2013 that engage in philosophical areas, such as ethics or metaphysics. He uses this to support alternative views of the philosophical canon; for example, he argues that, between 1968 and 1975, the most significant advances in philosophy were made by ethical theorists, such as Rawls and Single, whereas the traditional understanding of that period focuses on the philosophy of mind and language (Weatherson, 2020).

  As another notable example, scholars used embedding representations to evaluate a claim from Kimberlé Crenshaw in her development of the concept of intersectionality (Crenshaw, 1991). She argued that discrimination is not merely 'additive'— the discrimination that a Black women would face is distinct from the discrimination which a women would experience and the discrimination which any Black person would experience. The authors argued that this claim was supported from their finding that the vector representation of the terms 'Black' and 'woman' combined was distinct from the vector representation of 'Black woman.' This methodology, by itself, could be challenged (for example, the authors did not use disambiguation techniques, which would have more accurately mapped the concept 'black' to its racial connotation). But, as the authors aimed to illustrate, a computational analysis of embedding representations can be effectively woven into a wider philosophical argument to support innovative methods of analysis (Herbelot et al., 2012).

*2.5 Our Approach*

In the context of prior work, we contribute through our methodology as well as the scale of our analysis. Our methodology, outlined in the subsequent section, creates our reference network by treating authors as nodes and in-text references to other authors in their work as edges. We classify the context of each reference to measure the frequency with which authors are discussed within subdisciplines of philosophy. This approach carries an advantage in clearly delineating the impact of individual thinkers across a broader range of philosophical discourse. Our BERT-based topic classification also further demonstrates the effectiveness of advanced deep learning text classification through transformer architectures, which have become more prominent within the past few years (Wilkens et al., 2023).

The scope of our analysis covers an area that previous computational approaches have only studied in smaller subsets. This makes it, to our knowledge, the most comprehensive computational analysis of historical philosophical works to date. The range of our corpus—including ancient Greek thinkers, Christian theologians, modern philosophers, and influential scientists—is a large focus of scholarship and undergraduate pedagogy within both the academic discipline of philosophy, as well as the humanities broadly. The widespread recognition of our authors allows our findings to carry relevance for this broader academic community.

**Methodology**

*3.1 Data Selection*

We collected our dataset through the Project Gutenberg API (Gutendex, n.d.). Initially, we configured a Python script to download all available texts categorized as philosophy, written in or related to English, and available in .txt format, resulting in an initial corpus of 2,819 texts. To capture how philosophy has historically intersected with adjacent fields, we extended our script to download up to 200 texts from six additional categories: literature (218 texts), science (220), politics (215), religion (206), physics (157), and mathematics (54). In the case of mathematics and physics, the final counts reflect the total number of available texts. In the other categories, slight variation from the 200-text target resulted from batching and attempts to download works not available in the proper format. Our script also generated a spreadsheet containing five metadata fields for each text: its author(s), category, title, the author's birth year, and the author's death year.

      We chose Project Gutenberg due to its wide variety of canonical works with consistent formatting and accessible metadata for each book. Compared to the Internet Archive, Project Gutenberg provided a more defined scope by only providing works within the public domain (these works typically become available 70 years after an author's death). In Section 6, we discuss how our dataset can be further expanded through works published in other languages, works published by contemporary authors, and the inclusion of authors without surviving published works, such as the pre-Socratics. Its current range allowed us to maintain a focus on works that strongly aligned with our intellectual backgrounds while providing a comprehensive

analysis which expands previous efforts in applying computational approaches to historical works of philosophy.

*3.2 Data Preparation and Cleaning*

From our initial download of 3,889 texts, we filtered out duplicate works, works with unknown authors, and works in which Project Gutenberg did not have either the birth year or the death year of the author. Since our reference collection program, outlined in 3.3, relies on the direct string matching of author surnames[1], we manually eliminated authors who would generate too many false matches. Surnames such as 'Wake,' 'Bell,' and 'Post' would match with any instance of these words, and the scale of our dataset made context disambiguation impractical for each of these cases. While most major philosophers had distinct surnames that allowed for highly accurate string matching—such as 'Kant,' 'Rousseau,' and 'Plato'—a handful of surnames from important figures were ambiguous, such as Adam 'Smith,' Martin 'Luther,' or Francis *'Bacon.'* In these cases, we kept them in the original dataset for the reference collection program and applied context disambiguation later.

*3.3 Reference Collection and Validation*

After conducting these preparations, our final dataset contained 1,087 unique authors and 2,245 unique texts with a publication date which ranged from ~550 BCE to ~1940 AD. We then developed an additional Python script to identify and collect each time an author is mentioned in each text. This program iteratively scanned for each in-text reference of an author from each text to save three pieces of information: the author which is referenced, the author of the text, and 150 characters of surrounding context in which each reference occurs. To manage the computational costs, we excluded references from an author to themselves and limited the number of references from any single text to 250. This limit was particularly important for cases like biographical works—for example, George Grote's biography of Aristotle contains thousands of references to him, which would have skewed our analysis.

---

[1] With the exception of Greek philosophers, who did not have surnames.

After collecting these references, we manually inspected our dataset to validate its accuracy. For well-known authors, direct string matching consistently produced accurate results, as we could easily recognize references in context. For lesser-known figures, we reviewed multiple extracted references to each author and consulted secondary sources—primarily the Stanford Encyclopedia of Philosophy (Zalta et al., n.d.)—to confirm their validity. Based on this process, we constructed a "main dataset" consisting of authors for whom we manually validated several representative references and had high confidence in the precision of every detection. The full set of extracted references, including unvalidated authors, is referred to as the "expanded dataset."

We also identified 6-8% of references to authors which came from editors in introductions or footnotes. In our view, these cases could either be considered complementary to our analysis, as editor references indicate meaningful connections between authors, or could be considered noise in detecting true cases in which the author intentionally makes a reference in their original writing. To account for this, we created an additional subset of our data from our main dataset which filters out instances where an author references an author past their death. From our manual inspection, this filtered out at least 50% of cases where references were made by an editor[2]. In our analysis, we only used this subset of our data to validate patterns found in our main dataset to ensure that references made from editors did not skew our findings.

---

[2] There are also likely cases where this operation filtered out false references incorrectly identified with direct-string matching. However, we believe these cases negligible for two reasons: first, we were unable to identify any of these cases; second, this subset was derived from our main dataset, where we have a high-level of confidence on the accuracy of each reference.

After this process, we arrived at three subsets of our collected in-text references, each of which had different advantages for our analysis (see Table 1). While we will detail their differences in Section 4, our key findings will outline patterns common in all subsets of our data, unless explicitly noted otherwise.

*Table 1: Composition of the Main datasets*

| Dataset | Total Authors | Total References | Description |
| --- | --- | --- | --- |
| Main | 171 | 96,994 | Most important philosophers for computationally intensive analysis |
| Expanded | 1,087 | 294,970 | Largest breadth of authors included, only necessary filters applied |
| Filtered | 170 | 92,945 | Additional filters to exclude references made by editors |

*Section 3.4: Semantic Classification*

To more precisely explore how the influence of figures compares across fields, we applied topic modelling to classify each reference into eight predefined topics: ethics, politics, religion, mathematics, science, art, metaphysics, and epistemology. These topics were chosen to represent classically characterized subdisciplines of philosophy—metaphysics, epistemology, ethics—while also showing how philosophy has carried over into other fields—science, religion, and politics. We chose "art" in place of "aesthetics" as, from our testing, we found references classified as relating to 'art' more broadly captured how philosophy intersects with literary fields while also maintaining references which involve aesthetic discussions in philosophy. Likewise, "mathematics" was chosen in place of "logic" for similar reasons: we considered the ways philosophers have influenced mathematics to be of particular importance to explore, and found that references classified as mathematics would include discussions related to logic.

To apply topic modelling, we created a Python script that imported a pre-trained BERT model (Devlin et al., 2019) to create an embedding of each reference based on 150 characters of surrounding context (Sanh et al., 2019). We chose to use a BERT model as it is widely considered to be the state-of-the-art approach to generating embeddings, and recent work within the digital humanities has validated its usefulness in topic classification. We chose our particular model, *distilbert-base-uncased,* since it was designed for short snippets of text, excelled in inconsistent formatting, and was lightweight enough to be applied at this scale.

Due to its computational intensity, our script only assigned embeddings to our main dataset. We then used the model to generate embeddings of the eight terms used to represent our predefined topics and compared these to the embeddings of each reference through cosine similarity. As intended, references in contexts more closely related to the predefined topic achieved higher scores for that topic. We then set cosine similarity thresholds for each reference to classify them into our predefined categories. To maintain a wide breadth for analysis, we set our thresholds generously to include references only somewhat related to the predefined topic and minimize missing any references which should be included.

Through this approach, we derived eight subsets of our main dataset, each one corresponding exclusively to a predefined topic (see Table 2, Table 3). Since we did not impose a limit on how many topics a reference could be classified into, these eight subsets overlap with each other to varying degrees—often in proportion to how strongly the topics themselves are related. But, despite this, they each provided a distinct enough subset which allowed our analysis to more precisely characterize the relation between individual philosophers and their impact on the larger network.

*Table 2: Datasets Classified by Topic*

| Category | Total Authors | Total References | Example |
|---|---|---|---|
| Politics | 162 | 9,385 | John Dewey referencing Jeremy Bentham: "...through free competition and not by governmental favor. The stress that Bentham put on security tended to consecrate the legal institution of private prop..." |
| Art | 168 | 24,256 | Nietzsche referencing Schopenhauer: "...according to which, as according to some standard of value, Schopenhauer, too, still classifies the arts, the antithesis between the subjective and..." |
| Mathematics | 158 | 13,291 | Dante referencing Euclid: "...circle freely anything that is round, either a body or superfices; for, as Euclid says, the point is the beginning of Geometry, and, according to what he says, the point is the beginning of the circle..." |
| Religion | 167 | 21,681 | Volatire referencing John Milton: "...Leave it, then, to Milton to set Satan and Jesus constantly at war. Let it be his to cause a drove..." |
| Metaphysics | 168 | 26,025 | Hegel referencing Kant: "...ideas of space and time contain, and showed in them their contradiction; Kant's antinomies do no more than Zeno did here..." |
| Science | 170 | 45,479 | Bertrand Russell referencing Rene Descartes: "...the science of dynamics was rapidly developing in Descartes' time, and seemed to show that the motions of matter could be calculated..." |
| Ethics | 165 | 32,507 | Francis Bacon referencing Plato: "...no man can speak fair of courses sordid and base. And therefore, as Plato said elegantly, 'That virtue, if she could be seen, would move great love...'" |
| Epistemology | 165 | 46,844 | John Stuart Mill referencing Spinoza: "...here proved from two so-called axioms, equally gratuitous with itself; but Spinoza ever systematically consistent, pursued the doctrine to its inevitable..." |

*Section 3.5: Network Generation and Analysis*

We constructed an adjacency matrix from each subset of our data, where rows represent an author making a reference and columns represent an author receiving a reference. These graph representations gave us access to the same computational tools used to analyze citation networks derived from formal citations.

With Python's NetworkX library, we computed metrics to capture information about each individual philosopher (Hagberg, Swart, and S. Chult, 2008). These included citation counts (total number of incoming and outgoing references), unique connection counts (in-degree and out-degree centrality), and measures of betweenness centrality to analyze their network position.

We also applied the Louvain community mapping algorithm to analyze how philosophers are clustered based on their position in our network, and how these communities compare to the general understanding in how philosophical schools are divided (Blondel et al., 2008). Additionally, we measured the reciprocity scores of our network to find the frequency with which authors mutually reference each other. Due to the complexity of this project, we used the most standard, well-documented algorithms to compute each of these metrics to keep our results interpretable.

In Section 4, we applied these algorithms to analyze the general characteristics of our network's structure, and then later used them in Section 5 to supplement our analysis of specific philosophers as they relate to the wider network. We also identified countless other patterns which, for brevity, we could not include here. For this reason, we developed PhilBERT, a publicly available interactive tool for users to dynamically interact with our dataset and to illustrate our results.

*3.6 Visualization Tool*

We developed PhilBERT using the D3.js JavaScript library to create an interactive visualization tool to explore our reference networks (Bostock, Ogievetsky, and Heer, 2011). The tool's core design arranges authors on the x-axis, by birth year, and y-axis, by outgoing references, providing immediate visual context for influence patterns across time. Each author is represented by a node, with translucent bubbles scaling logarithmically with received references to indicate relative influence.

The interface offers several ways to explore philosophical influence. Users can adjust reference thresholds to focus on stronger connections, manually select specific philosophers for detailed analysis, or choose focal points to examine direct influence patterns. When a philosopher is selected as a focal point, the tool can highlight their most significant references and their most significant citing authors, enabling detailed exploration of influence chains. The tool supports multiple views of our data through eleven different matrices: our main dataset of 163 canonical philosophers, eight topic-specific networks (ethics, metaphysics, etc.), a filtered version removing potential noise, and our expanded dataset of all 1,088 authors. This flexibility allows scholars to examine both broad influence patterns and specific philosophical traditions.

**PhilBERT is publicly accessible at:** https://ogreowl.github.io/PhilBERT/

**Overall Network Characteristics**

*4.1 The Power Law Distribution of Influence*

The top two most referenced philosophers, Plato and Aristotle, receive around 20% of all references in our main dataset, and 10% of references in our expanded dataset (see Fig 1). The top ten most referenced authors receive ~40% of all overall references in our main dataset and ~20% in our expanded. While even people with a background-level knowledge would know of the central role of Plato and Aristotle in philosophy, our data indicates that their influence may even continue to be underestimated. As demonstrated in surveys (Szaszi et al., 2024), humans systematically underestimate concentrations in power law distributions, such as the American wealth distribution. From the current implications of our dataset, the distribution of references—and thus, by proxy, intellectual influence—may even be steeper.

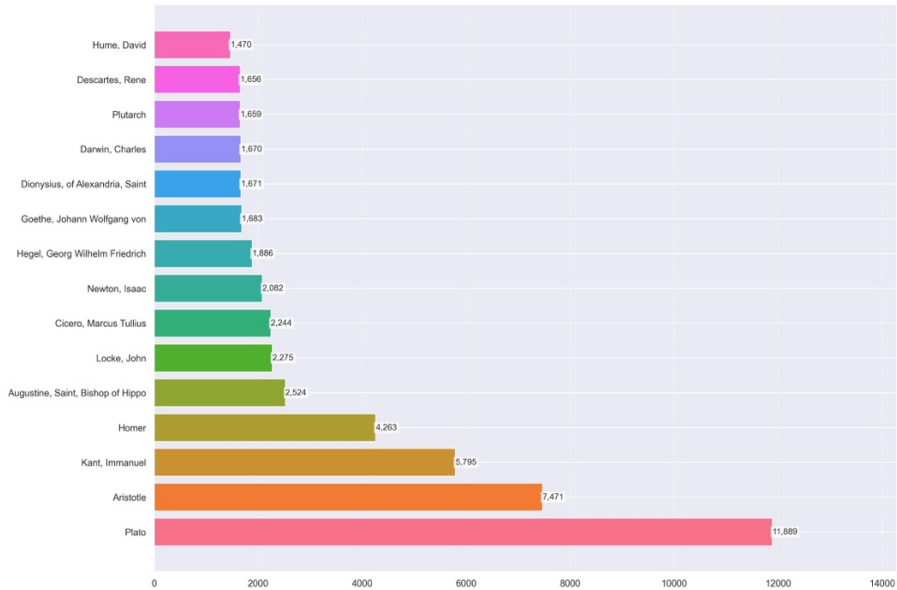

*Figure 1: Most Referenced Authors – Main Dataset*

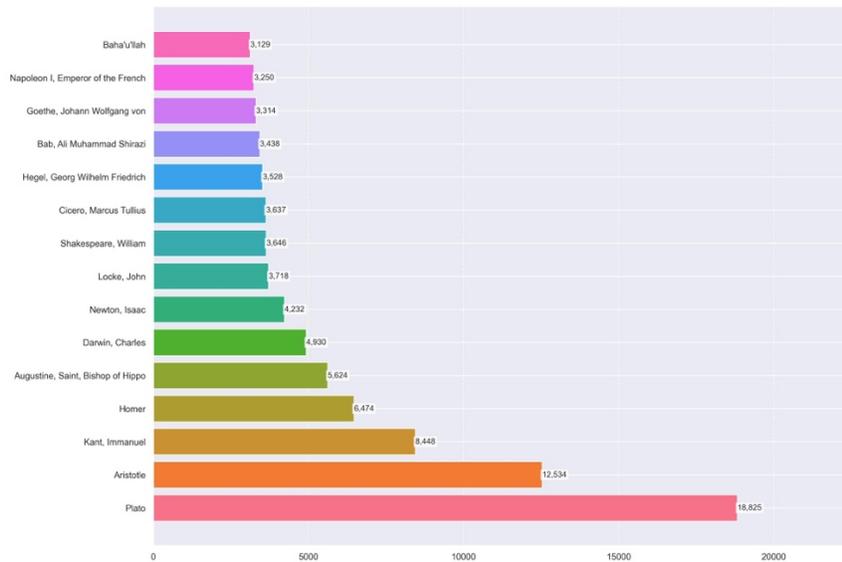

*Figure 2: Most Referenced Authors – Expanded Dataset*

When measuring in-degree centrality—total number of references from unique authors—the steepness decreases in our main dataset but increases in our expanded dataset (see Fig. 3, Fig. 4). In the main dataset, Plato scores 120 while the median stands at around 22; in the expanded dataset, Plato scores at around 480 while the median is less than 8. This indicates that the disproportionate influence of the most referenced figures increases in breadth as more authors are included—a relatively greater number of individual authors reference the top philosophers at least once, though relatively smaller number of authors engage with their works in depth.

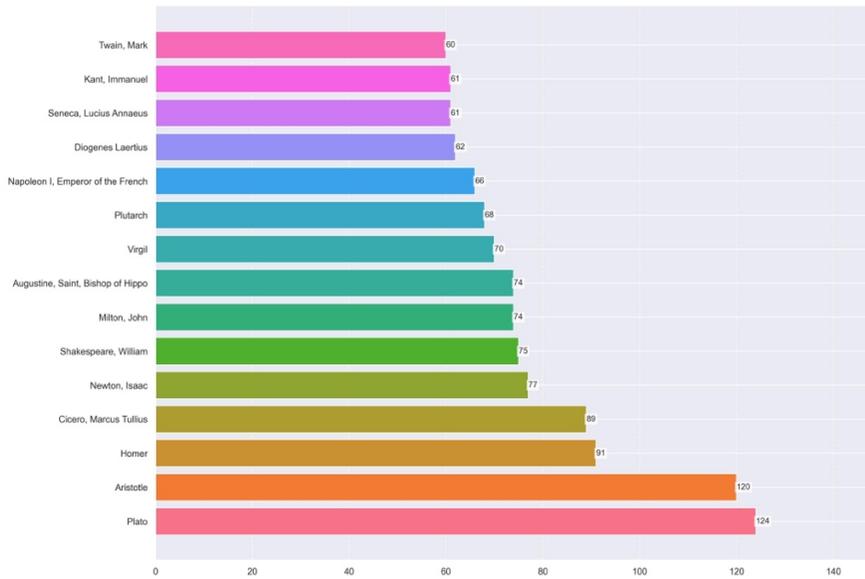

Figure 3: Highest In-Degree Centrality — Main Dataset

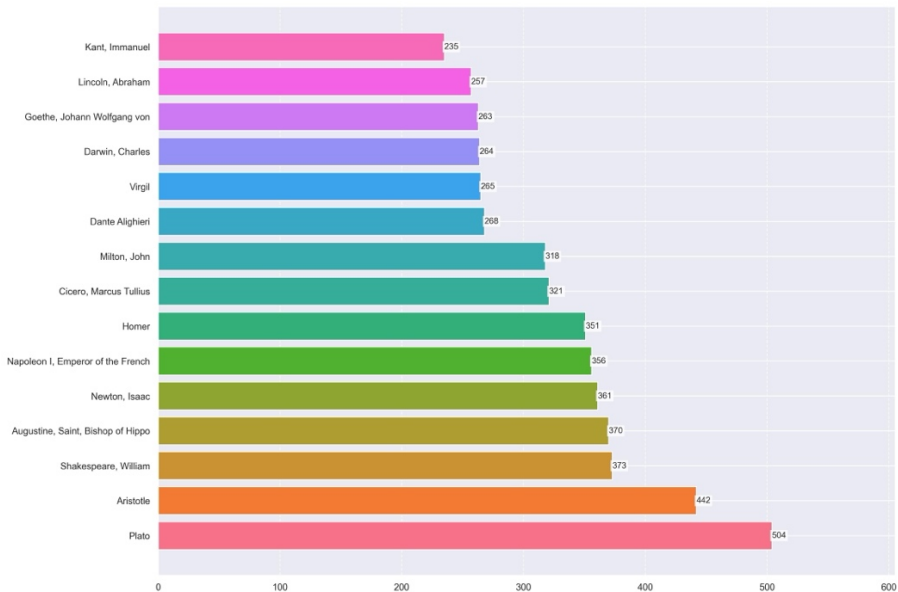

Figure 4: Highest In-Degree Centrality – Expanded Dataset

There is consistently a moderate correlation between earlier philosophers and higher centrality scores. The primary reason for this is somewhat intuitive—being around for longer allows earlier philosophers to receive more references throughout time. References to Karl Marx, for example, are rare in comparison to our understanding of his influence, which is likely because many references to him are made after the time range of our dataset. Another contributing factor to this is that, generally, only earlier philosophers with a higher level of importance are likely to have surviving publicly available works—the works of less influential philosophers which came earlier are more likely to be lost or poorly documented. To account for this, our visualization tool allows the user to filter out authors from distant time periods in order to directly evaluate an individual's relative influence with their contemporaries.

*4.2: Dominance Across Categories*

In our networks of references classified into specific categories, the most central thinkers remain dominant, with some expected variations among them (see Fig. 3). Kant, for example, scores closer to Aristotle in ethics, and surpasses him in epistemology and metaphysics. John Locke, Hegel, and Hume also receive proportionally higher references in these three categories together, reflecting how these topics were strongly interconnected in modern philosophical discussions. Saint Augustine and Saint Dionysius receive a greater percentage of references classified as religion; Darwin and Newton receive a greater percentage of references classified as science; and Shakespeare, Virgil, and Homer all receive a greater percentage of references classified as art.

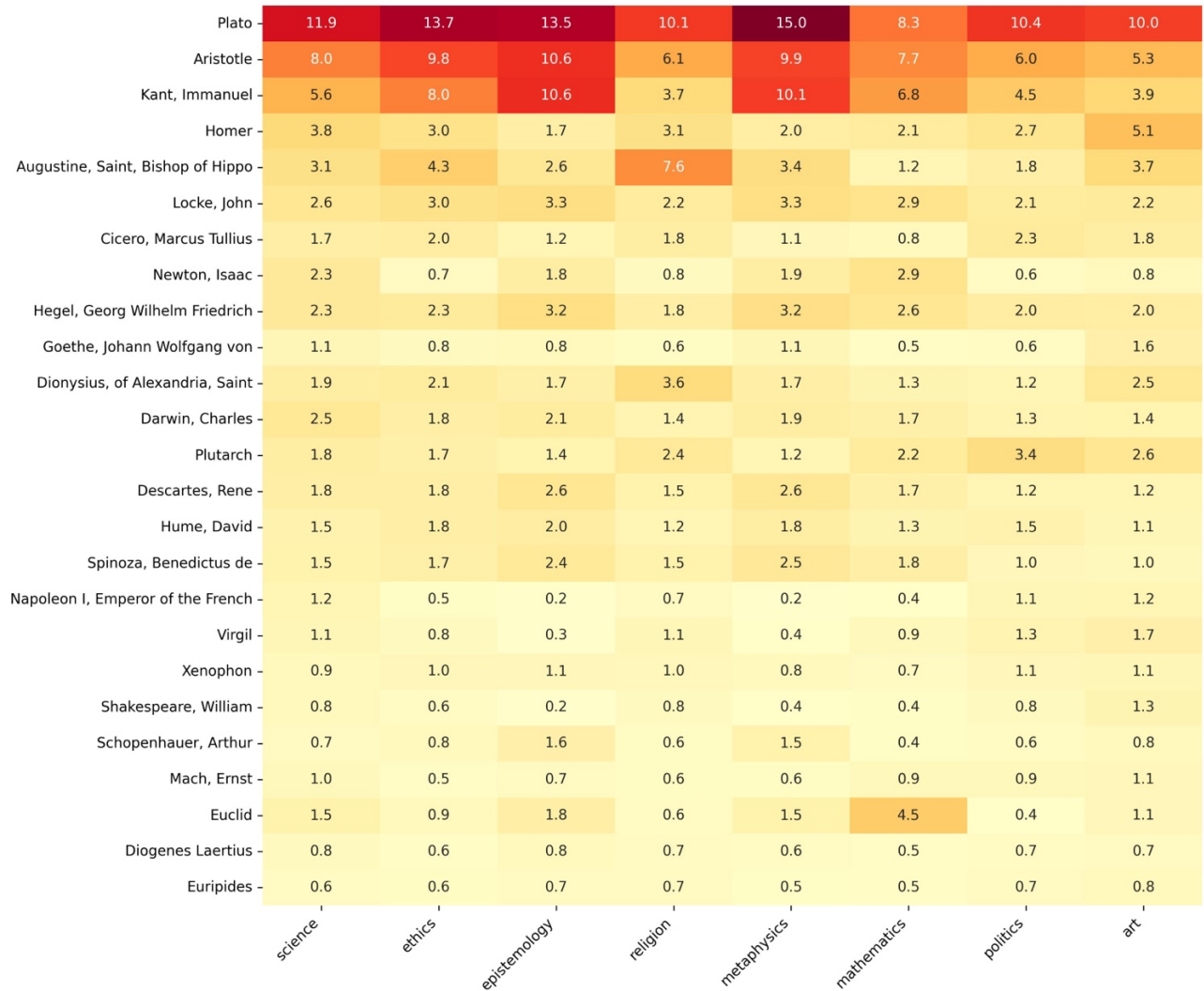

Figure 5: Percentage of Incoming References per Category

### 4.3 The Top-Down Spread of Influence

Our main network had a reciprocity score between 0.1-0.2, while our expanded network varied between 0.01-0.02, indicating that mutual references are only an occasional occurrence. Since the reciprocity score has a notable decrease when additional thinkers are included, philosophical influence appears to largely spread "top-down"—major thinkers impact many minor thinkers, but minor thinkers rarely impact major thinkers. It also implies that minor thinkers are far more likely to mention the most important philosophers than they are to mention less significant contemporaries.

One common exception to this is when a major thinker references a slightly less influential thinker, such as when Kant references George Berkeley. This is more common than when a highly influential figure references an obscure author. We hypothesize that one contributing factor to this phenomenon could be an 'immortalization' effect: if an extremely influential thinker commonly references someone, it draws attention to that person, making them less obscure. Such is the case of Nietzsche's frequent commentary on the works of Schopenhauer, which created a renewed interest in his works.

*4.4: Other Centrality Metrics*

Philosophers with the most outgoing references tended to be professors at European institutions in the latter half of the modern period (see Fig. 6). Hegel, Schopenhauer, Bertrand Russell, John Dewey, and Frazer all frequently make references to figures across our dataset, which indicates their role in preserving—and perhaps shaping—our understanding of the philosophical canon today. Out-degree centrality did not differ much from total outgoing references, with the exception of proportionally lower scores for biographers (Tyerman and Grote) whose works focused heavily on a handful of figures (see Fig. 7). Unlike for incoming references, there was a smoother distribution for measures of out-degree centrality.

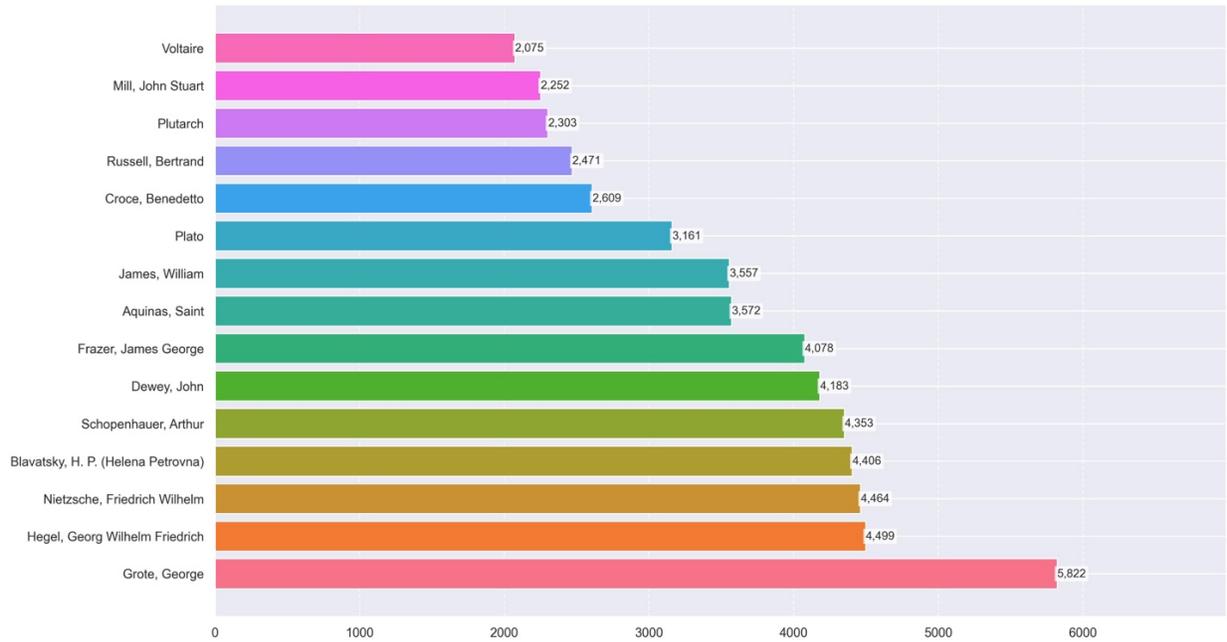

*Figure 6: Highest Number of Outgoing References*

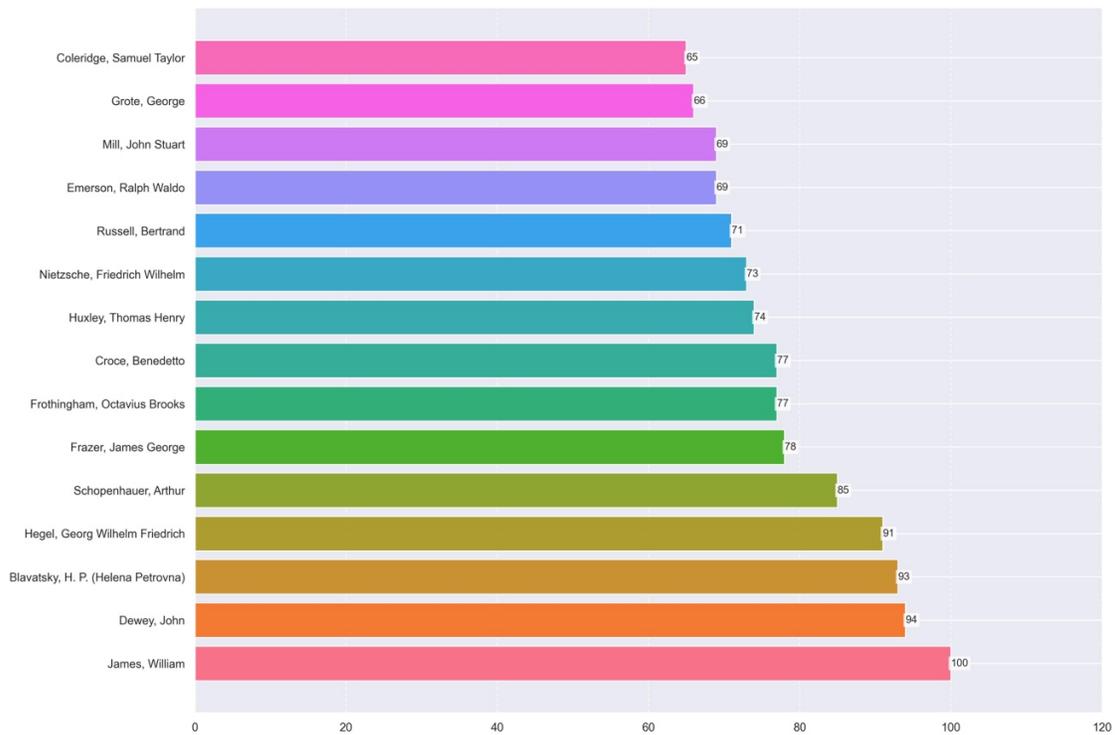

*Figure 7: Top Scoring Authors in Out-Degree Centrality*

Our measures of betweenness centrality, which measures how often nodes act as bridges between other nodes, often included historical figures who embody the archetype of a 'Renaissance Man'

— both in terms of literally living in the heart of the Renaissance and in terms of achieving expertise across a range of disciplines (see Fig. 8). Goethe, for example, was a mild influence in philosophy, but also had a strong influence in poetry, theatre, and even some scientific subjects, such as botany and colour. Likewise, Francis Bacon had a great influence in the development of scientific methods and through his career as a politician, both of which are in addition to his philosophical importance. Benjamin Franklin also fits this archetype quite well, having strong influence as an inventor, scientist, publisher, and, of course, as a founding father of the United States. The betweenness centrality of Mark Twain and Voltaire is explained by somewhat different reasons: they both popularized philosophical concepts through their literary works, creating widely cited novels outside of philosophy. The only top figures in betweenness centrality with reputations exclusively as philosophers are Spinoza and Hegel, which seems to indicate their impact on niche philosophical schools which branch off from the main philosophical canon.

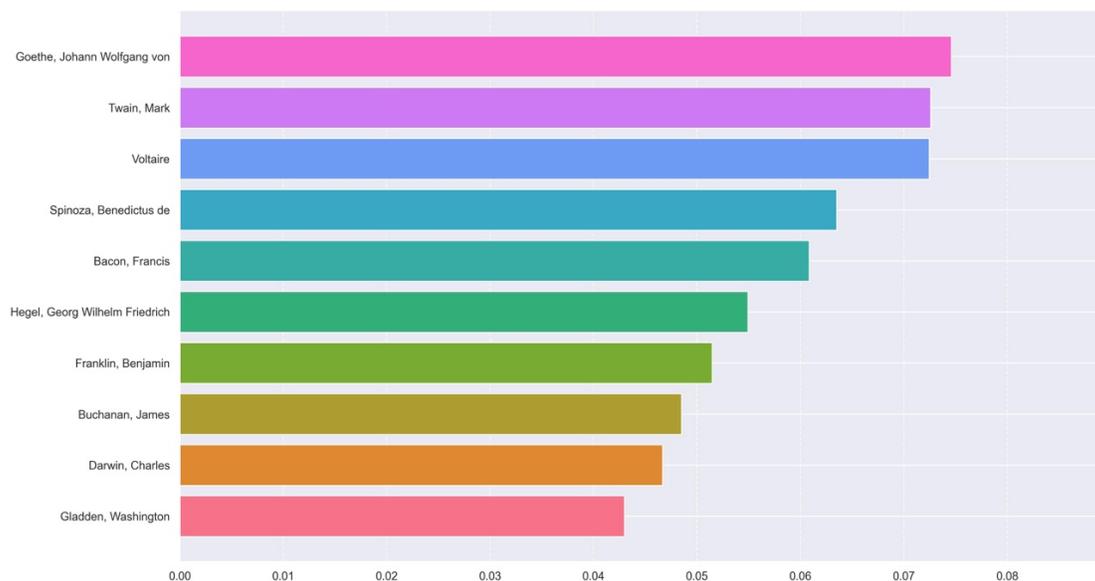

*Figure 8: Highest Scoring Authors in Betweenness Centrality*

*4.5 Grouping into Subcommunities*

We ran the Louvain community detection algorithm on our network which, by default, clustered philosophers into four major communities: one composed of ancient philosophers, without Plato and Aristotle (Homer, Plotinus, Aristophanes, Plutarch) alongside earlier Christian theologians who engaged with them (Saint Augustine, Saint Ambrose), (see Fig. 9); a second community composed of figures associated with a literary, political, or scientific canon, but not strongly associated to the philosophical canon (Lincoln, Einstein, Twain, Benjamin Franklin, Oscar Wilde) (see Fig. 10); a third community that maps onto the main thread of Modern European philosophy (Kant, Rousseau, Voltaire, Descartes) (see Fig. 11); and a fourth community composing mainly of post-modern philosophers, often associated with starting distinct schools of thought (Freud, John Dewey, Marx) (see Fig. 12). Plato and Aristotle would vary between these groups with small algorithmic changes, which implies their strong connection across all communities in the network.

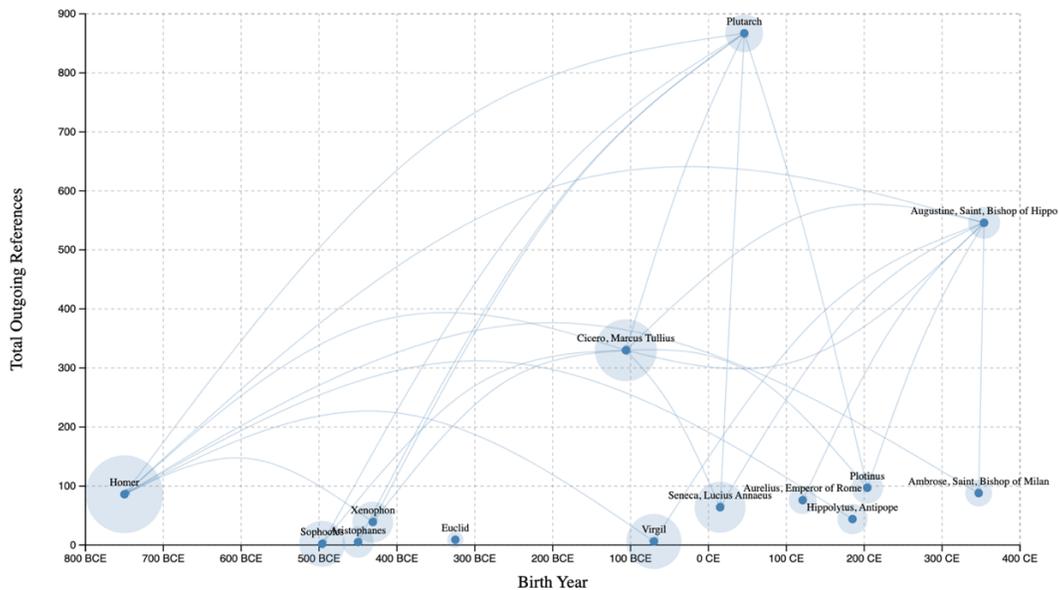

*Figure 9: Ancient Philosophers and Early Theologians*

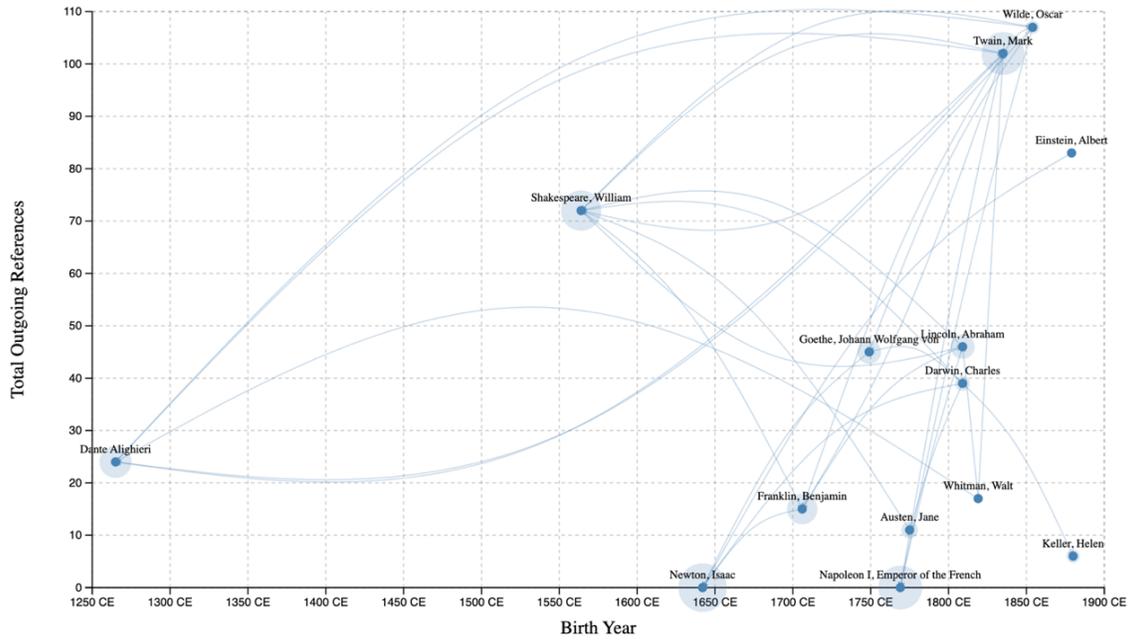

*Figure 10: Historical Figures Not Typically Categorized as Philosophers*

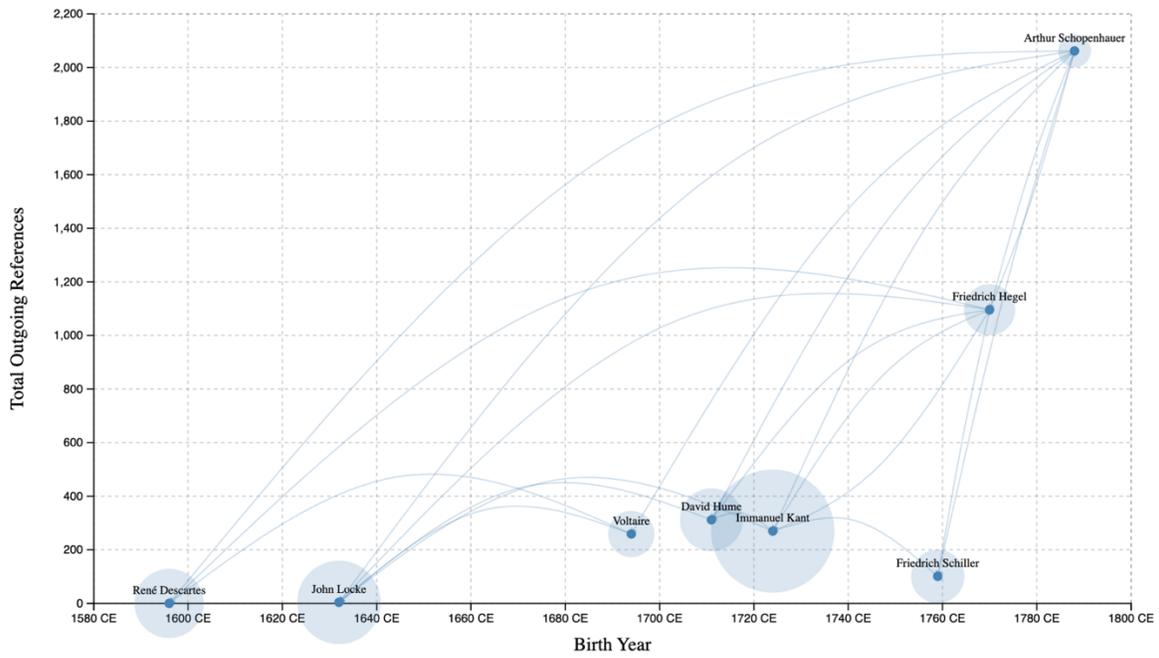

*Figure 11: The Modern Philosophical Canon Post-Renaissance*

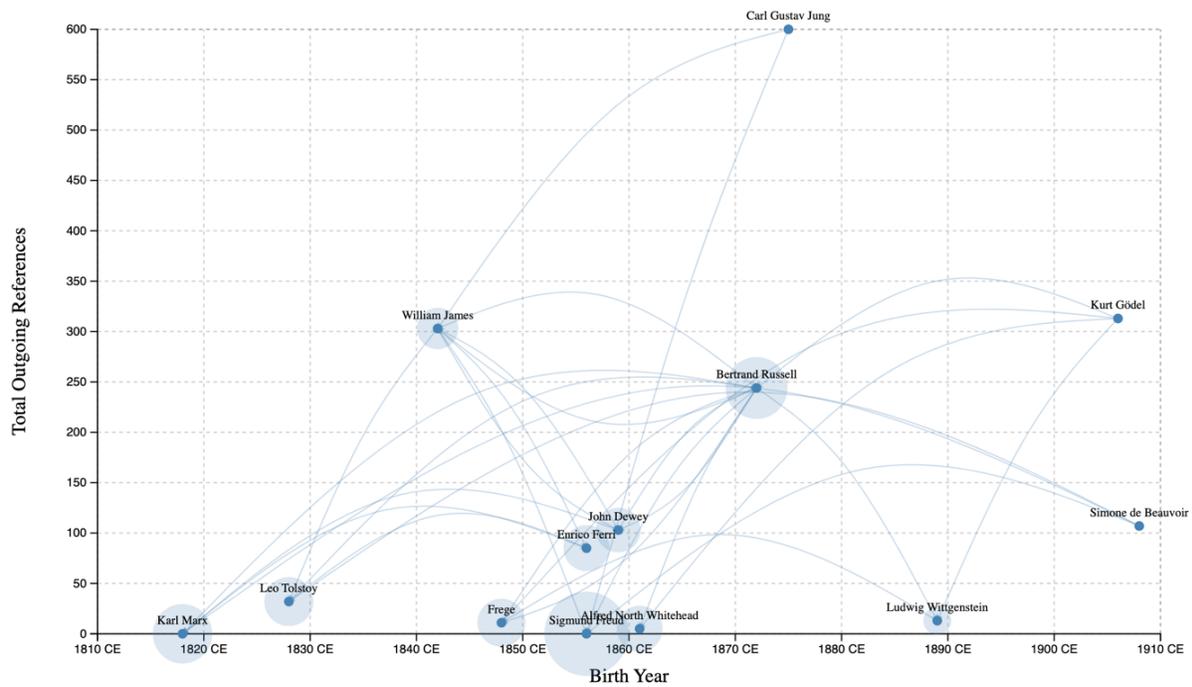

*Figure 12: Post-Modern Authors Associated with Pioneering New Fields*

When applied to the subsets of our network classified by topic, communities often contained more precise threads associated with well-defined philosophical debates. Our network with references classified as relating to politics grouped Rousseau, Mill, Hobbes, Bentham, and Adam Smith (see Fig. 13), which are all strongly associated with the development of modern political theory; our metaphysics network grouped Schopenhauer, Kant, Husserl, Spinoza, Hegel, and George Berkeley (see Fig. 14), whose ideas are frequently studied in comparison with each other. It is interesting to note how these communities did not map onto precise ideologies (ie, 'Marxists' or 'Kantian') but with collections of thinkers which often disagreed with each other. Dialogue appears to be a more accurate way to characterize philosophical lineages than agreement.

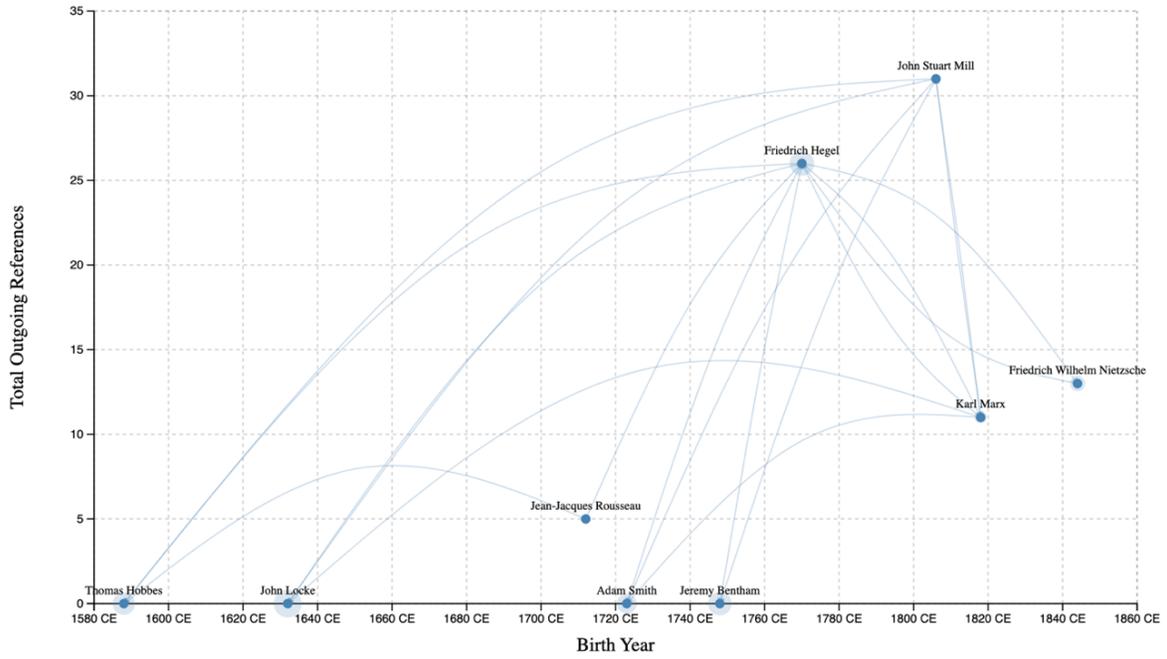

*Figure 13: Political Developments in Modern Philosophy*

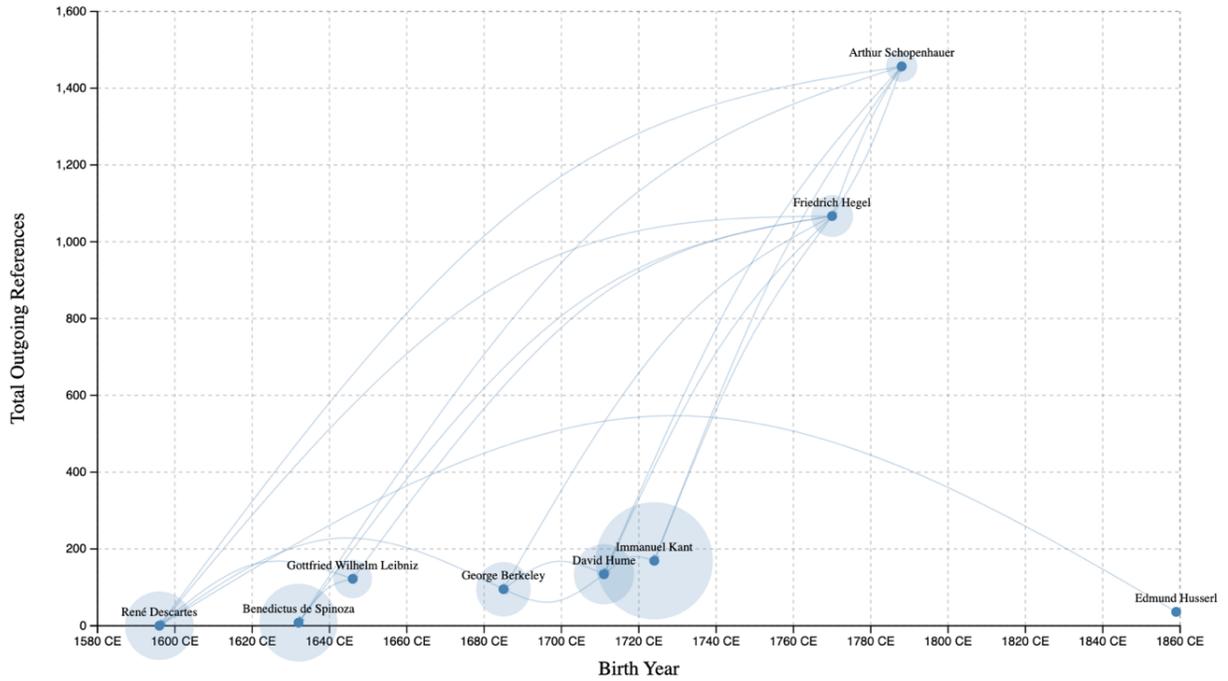

*Figure 14: Metaphysical Debates in Modern Philosophy*

The modularity score, which measures how distinct these communities are, ranged between 0.1 and 0.2. This is low enough to indicate that each of these communities are strongly interconnected, but high enough to indicate meaningful patterns in how various groups are positioned in our network which can be quantifiably identified.

**Analysis of Individual Authors**

*5.1 The Footnotes of Plato*

Alfred North Whitehead's claim that "Western philosophy is a series of footnotes to Plato" appears well-founded: he receives the most incoming references and highest centrality scores across all variations and measures of our network (Whitehead 1912). The vast majority of influential thinkers—in and beyond philosophy—mention him, and those who more frequently reference him are more likely to be influential themselves.

When focusing in on his reference network, we see the two major eras where figures reference him the most: Ancient philosophers—Aristotle, Plutarch, Origen, Cicero—and modern philosophers—Kant, Hegel, and Schopenhauer (see Fig. 15). Bertrand Russell describes the philosophical winter between these two eras as the "period of darkness", beginning after the barbarian invasion of Rome and lasting until the Renaissance (Russell, 1945). Interestingly, across this time period, the Byzantine Empire preserved ancient philosophical works, but did not play an important role in carrying it to the Renaissance thinkers. In fact, Islamic theologians Avicenna and Averroes as well as the Jewish philosopher Maimonides were the most active links during this time period (see Fig. 16). Among others, two main historical facts explain this: first, the Church heavily policed philosophy around this time, often punishing ideas which questioned Christian doctrine (such as in the case of John Italus). Secondly, many manuscripts were lost in the Fall of Constantinople in 1453 (Treadgold, 1997).

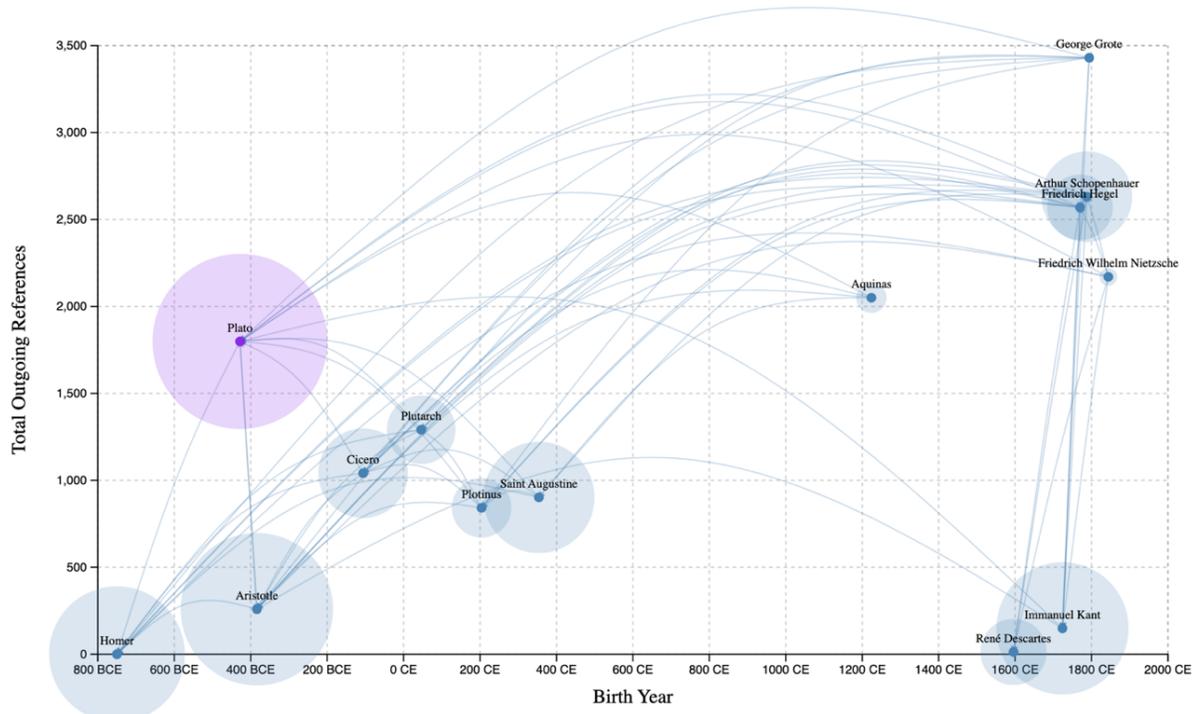

*Figure 15: Plato's Reference Network*

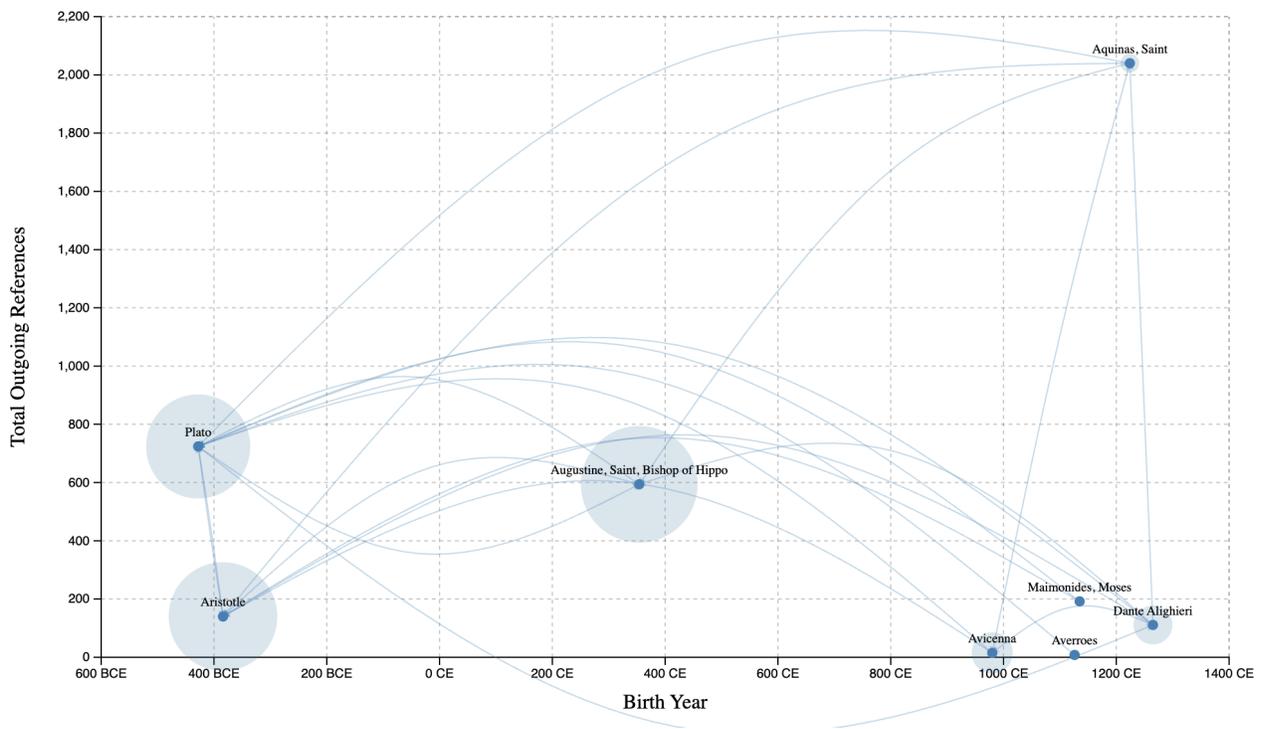

*Figure 16: Links During the Philosophical Winter*

Plato most frequently references his contemporaries (Sophocles, Aristophanes, and Euripides), and, while not featured in our analysis, the pre-Socratics. Homer is the only prominent figure in our dataset that Plato references who does not reference Plato back. It is quite possible that Homer's performance on our metrics—often achieving the fourth highest centrality—can be partially attributed to Aristotle's and Plato's frequent references to him (see Fig. 1, Fig. 17). His poems were the canonical artistic works of Ancient Greece, a time where cultural touchstones stayed far more constant than today. It is possible that, by attracting intellectual scholarship around Ancient Greece, Plato and Aristotle indirectly helped carry Homer into the modern canon.

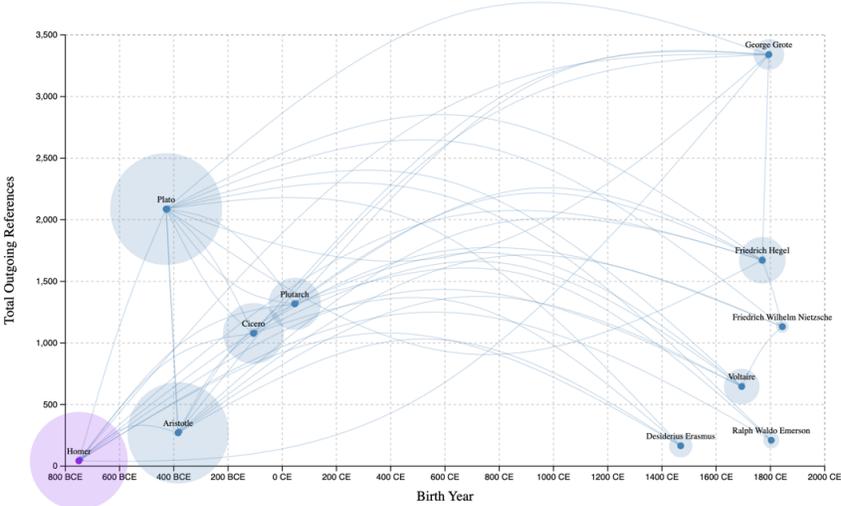

Figure 17: Homer's Reference Network

But while impressive, it is also likely that references to Homer do not reflect his level of influence as much as references indicate the importance of Plato and Aristotle. His works are more heavily engaged with in literary and artistic analysis (see Fig. 18); even when he is cited by scientific or philosophical thinkers, it is primarily for the purpose of allusion, metaphor, or insight into the values of the ancient world. The works of Aristotle and Plato were, in contrast, directly engaged with for their philosophical ideas themselves, often catalyzing competing interpretations of their ideas.

Just as Plato's work stands to Homer, Aristotle's work is more analytical and less literary than Plato's. Plato's dialogues asked many of the most fundamental philosophical questions, but only offered his true position by implication. Aristotle, in contrast, systematically explored various subject-matters explicitly, and developed more formal theories in logic, metaphysics, and ethics. Though even so, Aristotle still massively underperforms Plato across centrality metrics, often falling further below other authors in categories such as religion and metaphysics (see Fig. 5). Unlike Plato, there are historical gaps—such as in later antiquity and the early development of the Church—where Aristotle's work received less attention (see Fig. 18, Fig. 19). Aristotle receives almost as many references as Plato in mathematics—reflecting Euclid's engagement of his work, and possibly indicating Aristotle's proportionally greater importance in formal subjects. However, Plato's mathematical theory of numbers has started to receive more importance in the mid-20th century, in part due to Kurt Gödel (Kennedy 202) (see Fig. 20), which could squander Aristotle's last chance—and by extension, any other philosopher—in overcoming Plato's shadow in a major subject.

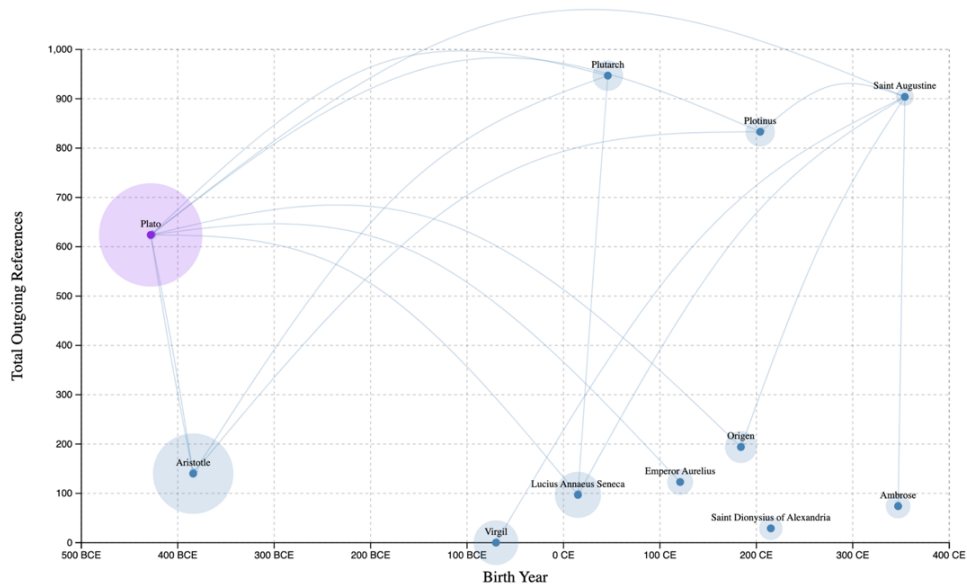

*Figure 18: Areas of the Network where Aristotle Underperforms*

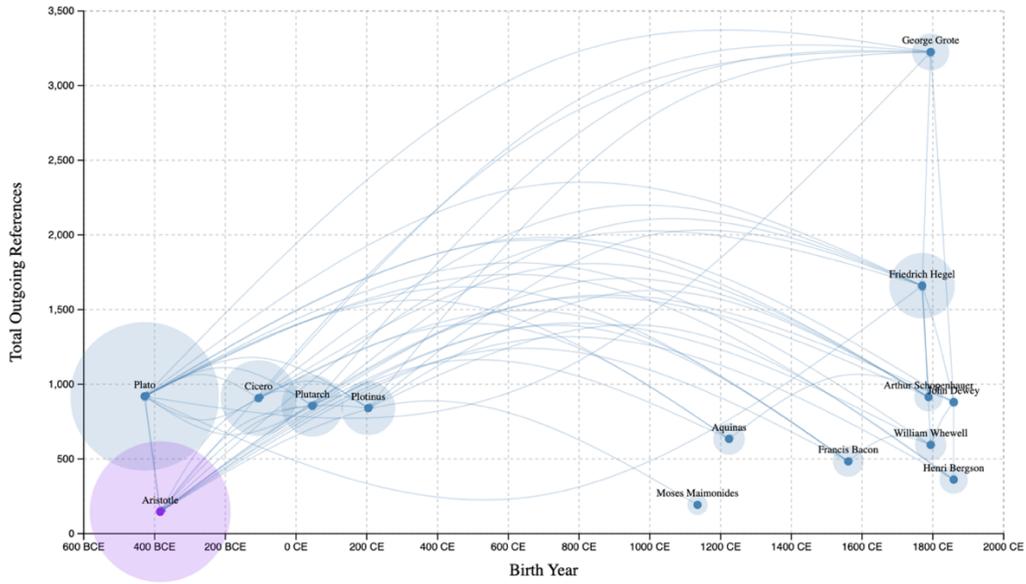

*Figure 19: Areas of the Network where Aristotle Underperforms*

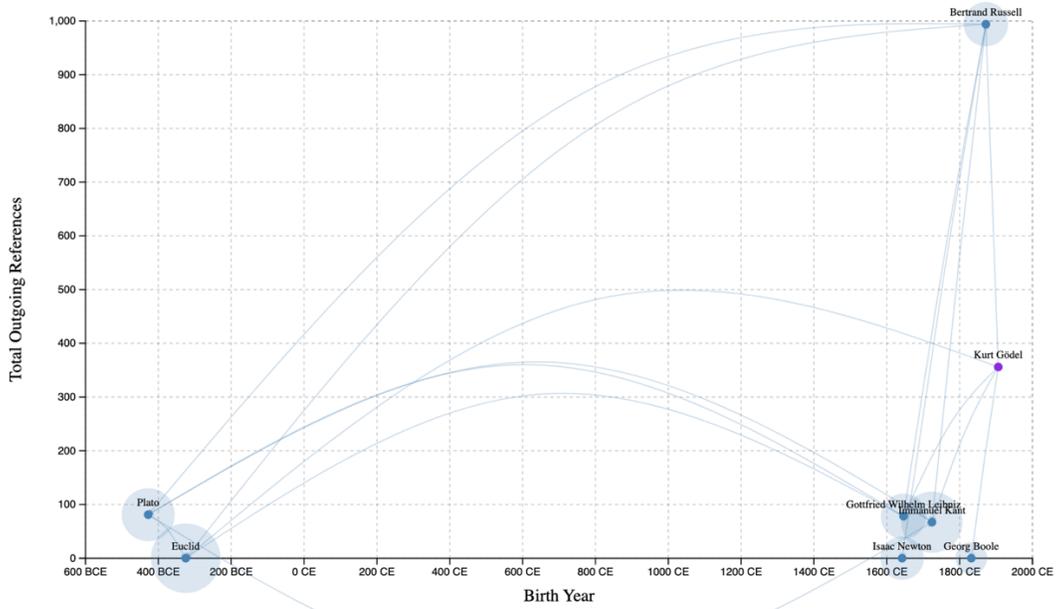

*Figure 20: Kurt Gödel's Outgoing References*

*5.2 The Spread of the Gospel*

The bridge between the ancient and modern eras of philosophy are Christian Saints and theologians. Saint Augustine, the fifth most referenced figure in our dataset, lived in 400 CE, around 800 years after Plato (see Fig. 21). He established principles for interpreting scriptures and argued for many ideas that have become core tenets of Christianity, such as the idea of the 'original sin' (Cary, 2008). He, along with his mentor Saint Ambrose, the Bishop of Milan, wrote about philosophers of Ancient Greece and Rome, incorporating ideals from Platonism and stoicism into their systematic development of Christian theology.

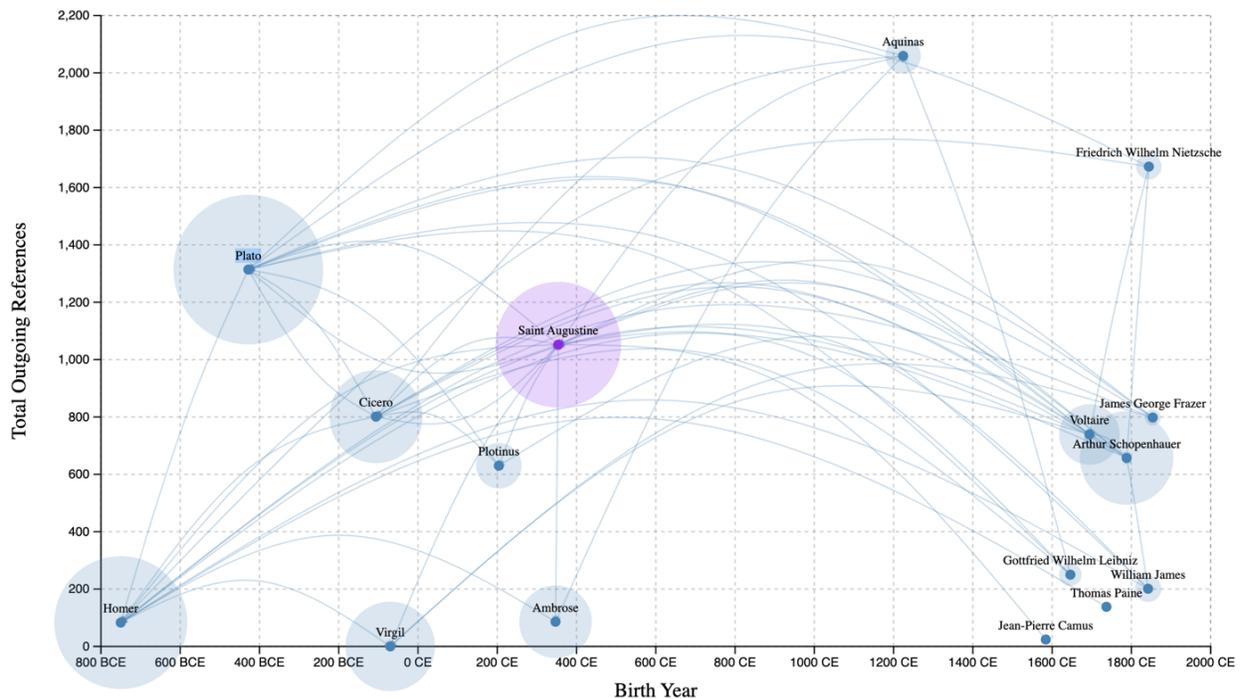

*Figure 21: Saint Augustine's Position in the Reference Network*

Notably, their references to Aristotle are sparse. Saint Thomas Aquinas, in 1200 CE, commonly references Saint Ambrose, Saint Augustine, and Aristotle, which aligns with his reputation as a synthesizer between Aristotelian and Christian ideals (Colish, 2009) (see Fig. 22). He is then later referenced by figures like Descartes, Montaigne, and Leibniz, becoming a core member himself and helping revive Aristotle as a central part of modern philosophy.

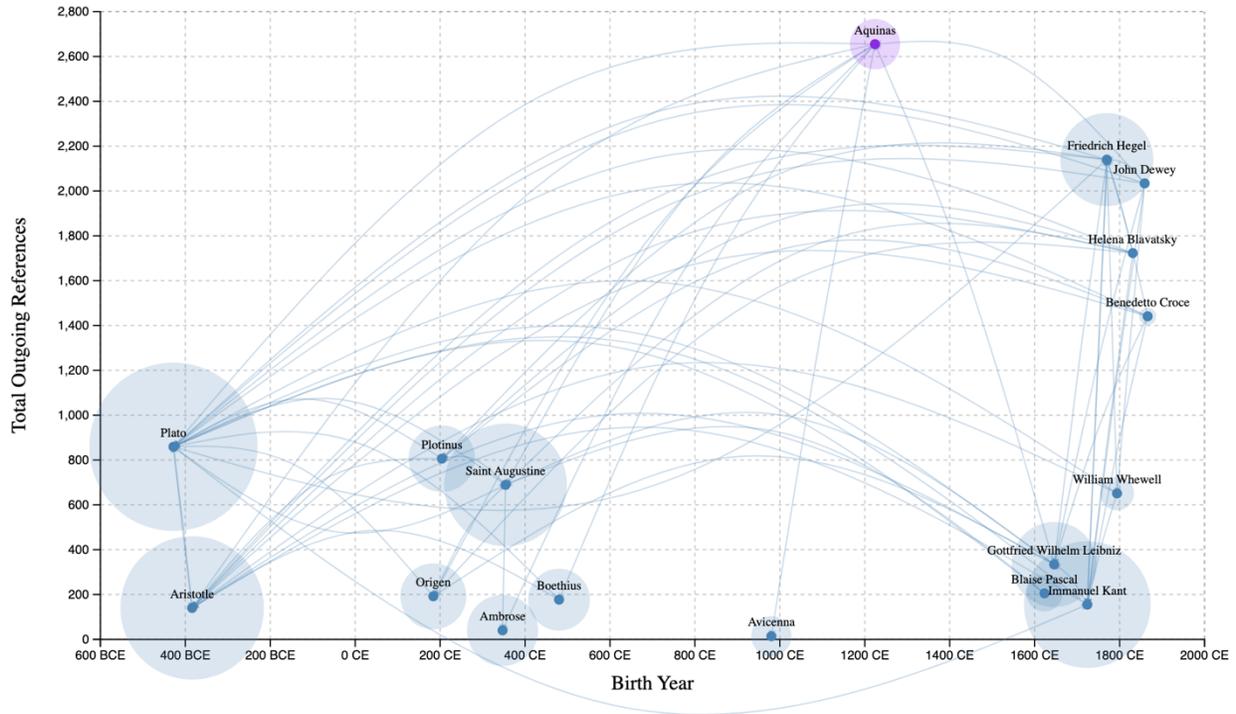

*Figure 22: Saint Thomas Aquinas's Position in the Reference Network*

After this bridge, the next major Christian theologian to have prominent influence in philosophy was Martin Luther (see Fig. 23). He took advantage of advancements made in the printing press, which were largely credited to Johannes Gutenberg, the namesake of our dataset. Around 1517, Luther began sharing short pamphlets and writings on his critiques of the Catholic Church, eventually giving rise to Lutherism in the reformation (Hendrix, 2015). Many of the most influential modern philosophers who came later were raised under denominations of Lutherism, including Kant, Hegel, and Nietzsche. However, these philosophers rarely referenced Luther in their major works. Luther receives a wide breadth of references from minor authors, but it appears that nature of his work made it unattractive for deep engagement and analysis.

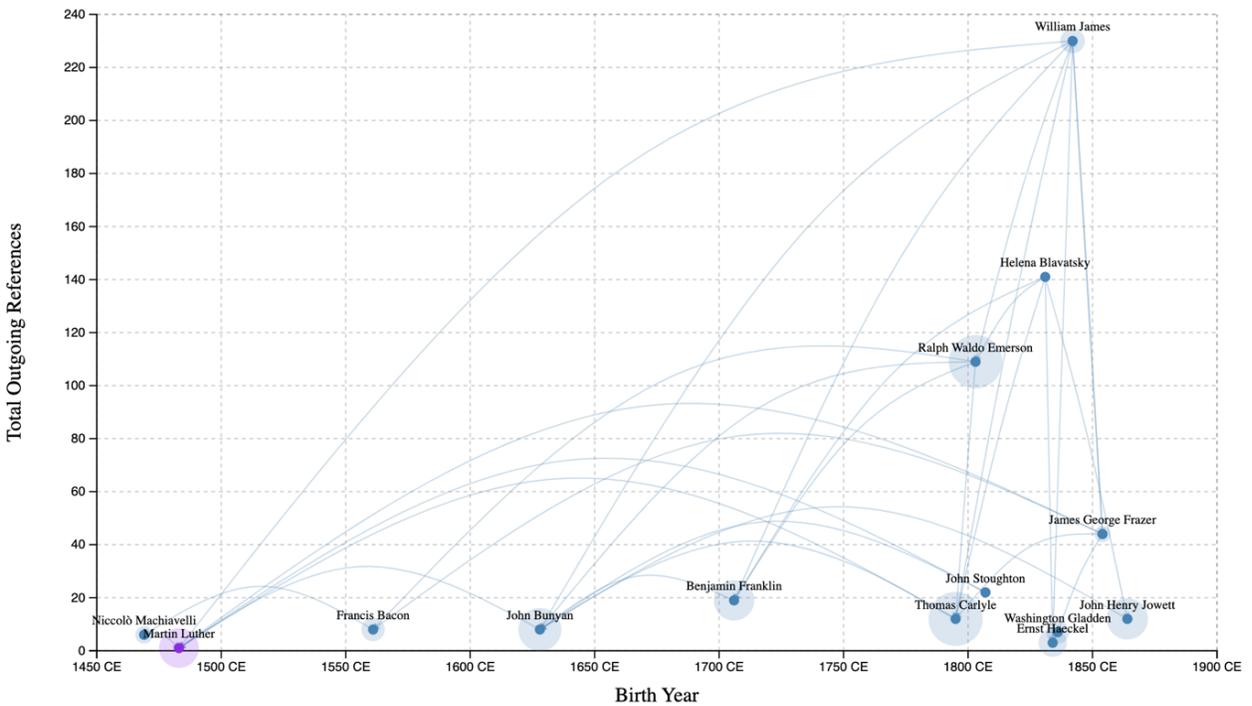

*Figure 23: Martin Luther's Position in the Reference Network*

## 5.3 The Copernican Shift of Kant

After Plato and Aristotle, the next most influential thinker in philosophy is Immanuel Kant (see Fig. 24). Despite remaining highly inaccessible to casual readers, his work has continued to be a mandatory touchstone among future philosophers in metaphysics and epistemology. His in-degree centrality is moderately less impressive than his total incoming references, which suggests a higher level of engagement with his work from a fewer number of thinkers (see Fig. 1, Fig. 3). Hegel and Schopenhauer, in particular, created entire separate threads of philosophy over their competing approaches to interpreting and building off of Kant. Hegel, then, most directly influenced Marx, which gave rise to one of the most impactful political ideologies in the 20th century; Schopenhauer, in turn, very directly influenced Nietzsche. Kant also influenced numerous thinkers who gave rise to distinct domains of study from philosophy, including influential psychologists (Freud, William James, Jung), mathematicians and logicians (Gödel,

Russell, Einstein), educational theory (John Dewey) and even feminist theory (Simone de Beauvoir). This is all in addition to Kantian ethics, which has, along with Utilitarianism and Virtue Ethics, become a primary ethical theory in popular discourse.

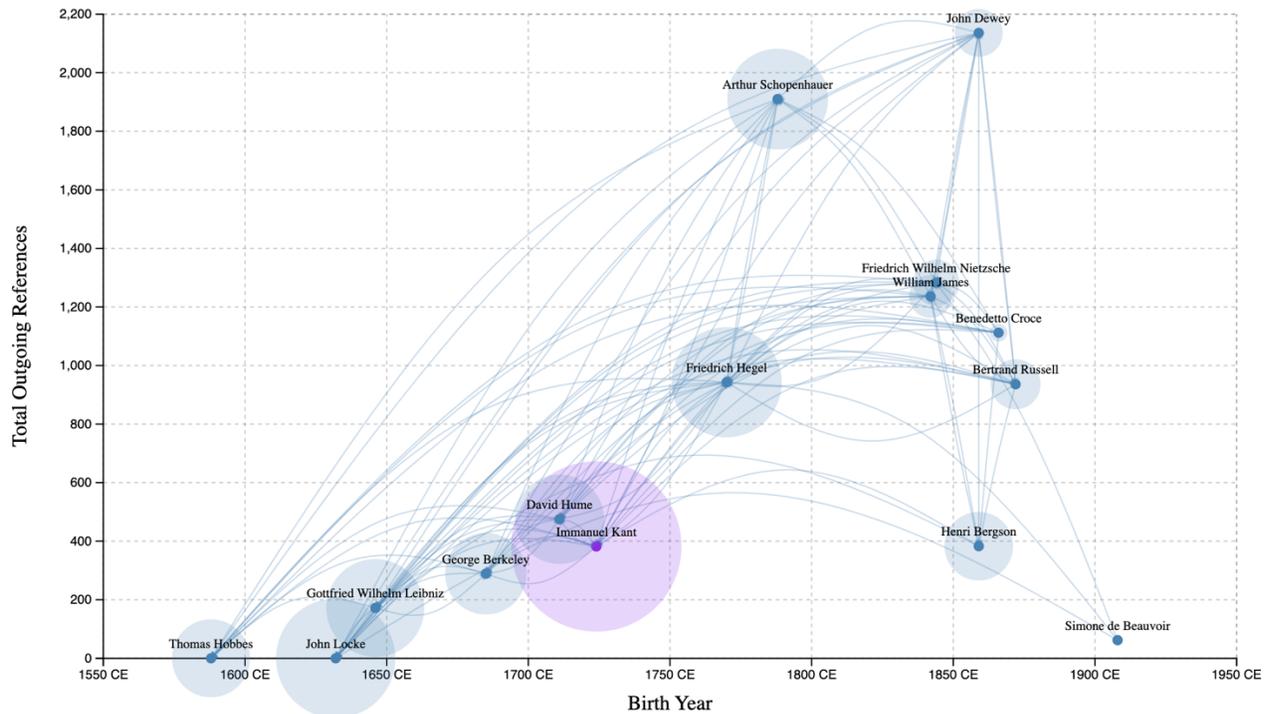

*Figure 24: Kant's Position in the Reference Network*

Outgoing references from Kant reflect his efforts in creating the most rigorously-developed response to the modern metaphysical debates—a thread starting from Descartes and then carried to Leibniz and then to Hume. He also frequently incorporated the theories of Newton, Euclid, and Galileo in his philosophical arguments, reflecting the fact that his "early works are more concerned with science than with philosophy" (Russell 1945, 678). Kant famously compared his ideas as being a "Copernican Revolution" in philosophy, in terms of offering an entirely new way to interpret the world than the methods of his predecessors (Kant, 1781/1998). We can debate as to whether his ideas in-themselves are worthy of that title, but there is little room to dispute the magnitude of his impact.

*5.4 Empiricism's Declaration of Independence*

In philosophy, Kant attempted to synthesize rationalist and empirical ideas, but his success "…cannot be admitted, at least from a historical point of view, for the followers of Kant were in the Cartesian, not the Lockean, tradition" (Russell). Thus, while Kant was able to bring the ideas of empiricist philosophers to his followers, the empiricist lineage continues to develop independently. This most definitively begins with Locke, who was inspired by Francis Bacon's success in grounding scientific discovery in empiricism as well as John Milton's ideals regarding human liberty (see Fig. 25). It then carries on to David Hume and George Berkeley, but also to thinkers such as Jean-Jacques Rousseau, John Stuart Mill, Adam Smith, and Thomas Paine.

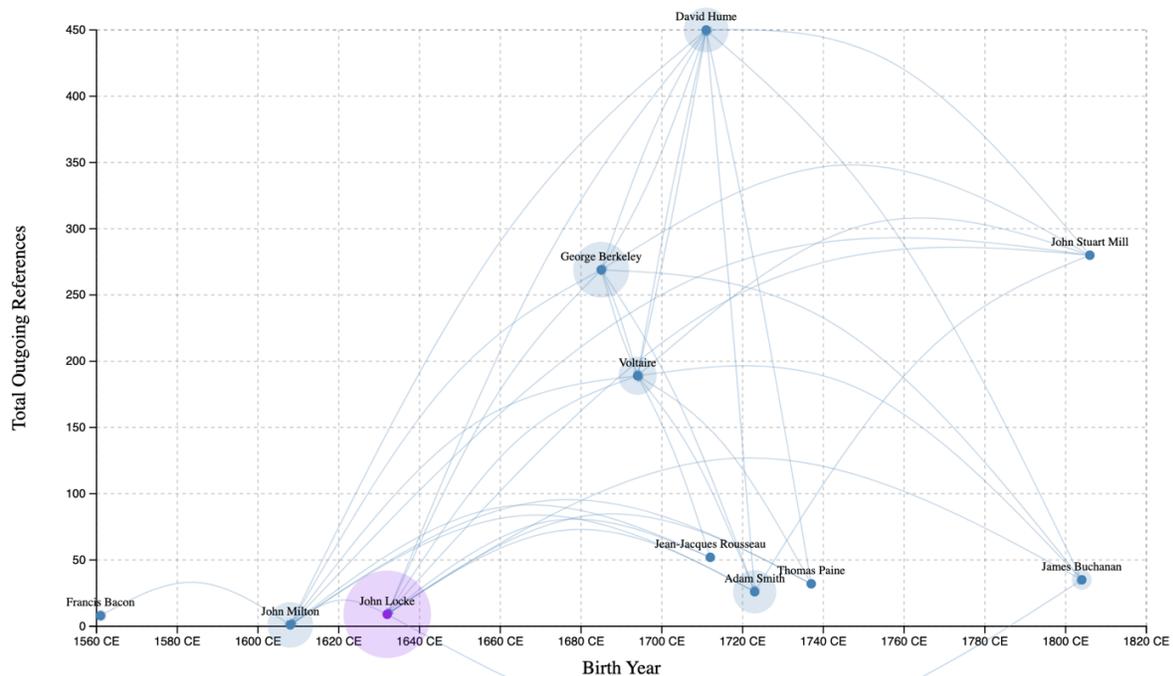

*Figure 25: John Locke's Position in the Reference Network*

This strand played a much more dominant role in shaping the political ideals of America: Thomas Jefferson references Berkeley, Rousseau, Thomas Paine, Smith, and Milton (see Fig. 26) while Benjamin Franklin most frequently references Locke and Milton, and was also known to be close friends with Hume (see Fig. 27). In the Declaration of Independence, Jefferson adapts

Locke's "Life, Liberty, and Property" into "Life, Liberty, and the Pursuit of Happiness", which has become a textbook example to illustrate the Enlightenment's impact on America's founding ideals. What is less known is how philosophy influenced Franklin's approach to revising Jefferson's work. In his original draft, Jefferson writes "We hold these truths to be sacred and undeniable…" Franklin, inspired by the Enlightenment's emphasis on reason and concrete pursuit of knowledge, suggests an edit this to the line which to what is now canonical to American history: "We hold these truths to be self-evident, that all men are created equal, that they are endowed by their Creator with certain unalienable Rights, that among these are Life, Liberty and the pursuit of Happiness" (United States, 1776).

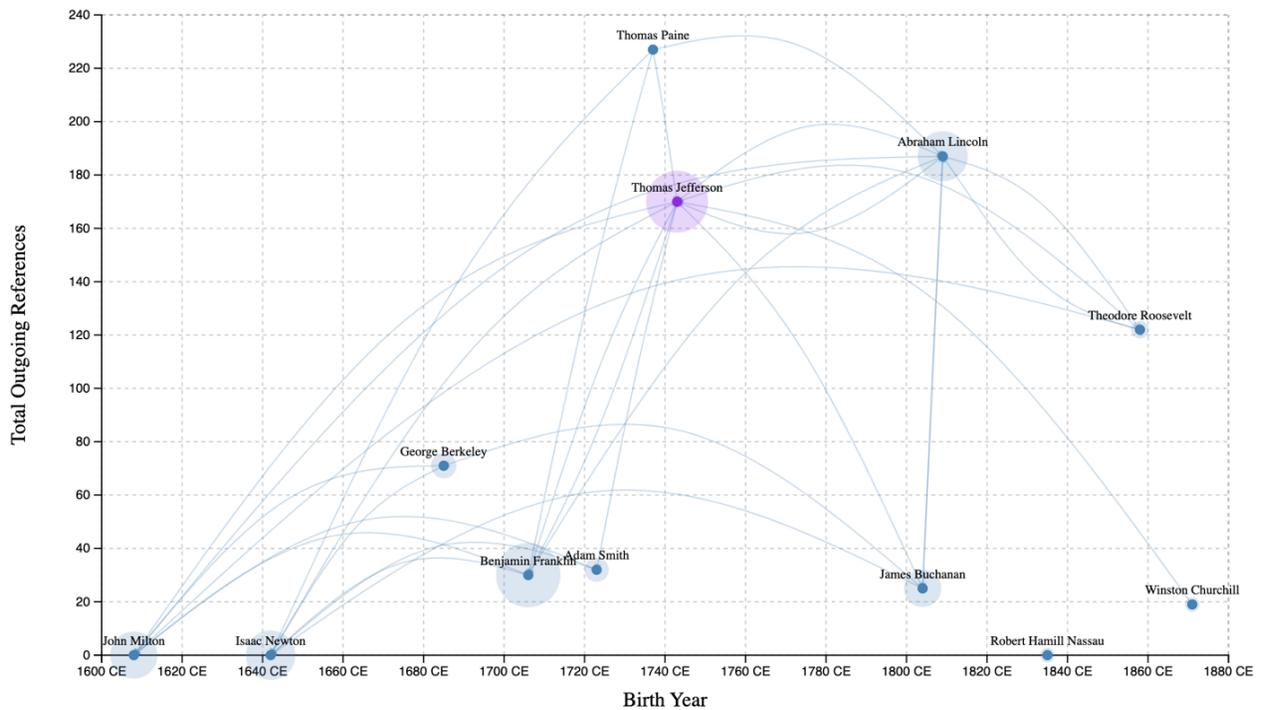

*Figure 26: Thomas Jefferson's Position in the Reference Network*

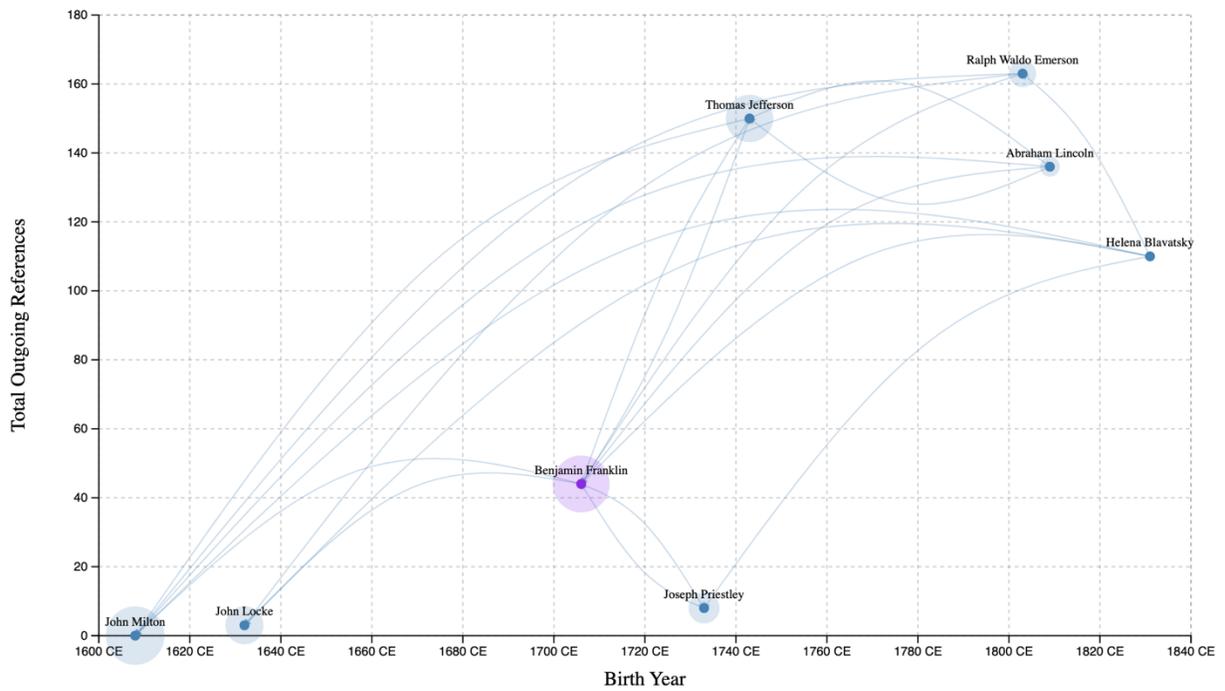

*Figure 27: Benjamin Franklin's Position in the Reference Network*

Russell attributed the dominance of empirical philosophy in England to the "victory of the Newtonian cosmogony" which "diminished men's respect for Descartes and increased their respect for England" (Russell 1945, 617). Newton also harmed Leibniz's reputation by accusing him of plagiarizing his discovery of calculus, creating a similar divide with British mathematics and the rest of Europe. Voltaire frequently references Locke in our network, and was shown to share British attitudes toward rationalist philosophers through his satire of Leibniz (see Fig. 28). This supports his reputation as the "chief transmitter of English influence to France" (Russell 1945, 672). Russell also credits the French Revolution for shifting British philosophy to focus more on politics, thus leading to their downstream influence in America.

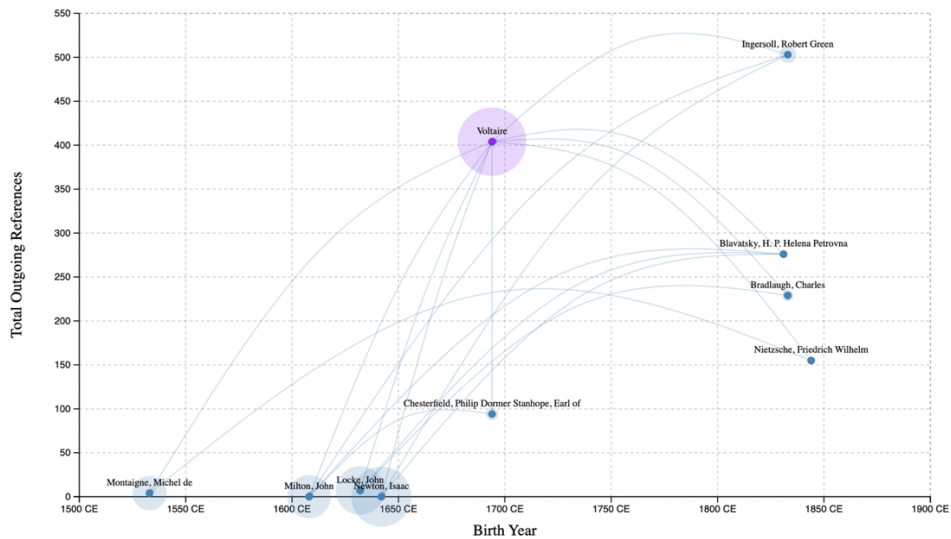

*Figure 28: Voltaire's Position in the Reference Network*

While Kant references Rousseau in his political essays, these works remained unimpactful. His political influence is only an indirect result of his influence on Hegel, who influenced Marx, and Schopenhauer, who influenced Nietzsche (see Fig. 29). Curiously, however, Nietzsche heavily disliked Kant, and Marx rarely references him. And while the Hegelian Dialectic informed the abstract foundation of his beliefs, Marx he most often cites the political works of the Empiricists, such as Locke, Smith, Berkeley, Hume, and Mill (see Fig. 30). Marx considered these thinkers to have pushed political progress forward but criticized them for being limited by the class structure which shaped their thinking. From Marx's view, Locke's emphasis on private property served to expand human rights during his time when the principle gave individuals power over feudalistic governments, but later served to reinforce injustices inherent to capitalism, which would need to be overcome by a communist revolution. These lineages characterized the development of political thought in Europe, from "through Bentham, Ricardo, and Marx, by logical stages into Stalin…" and "through Fichte, Byron, Carlyle, and Nietzsche, into Hitler." In America, "the ideas of liberalism have undergone no part of this development… where they remain to this day as in Locke" (Russell 1945, 618).

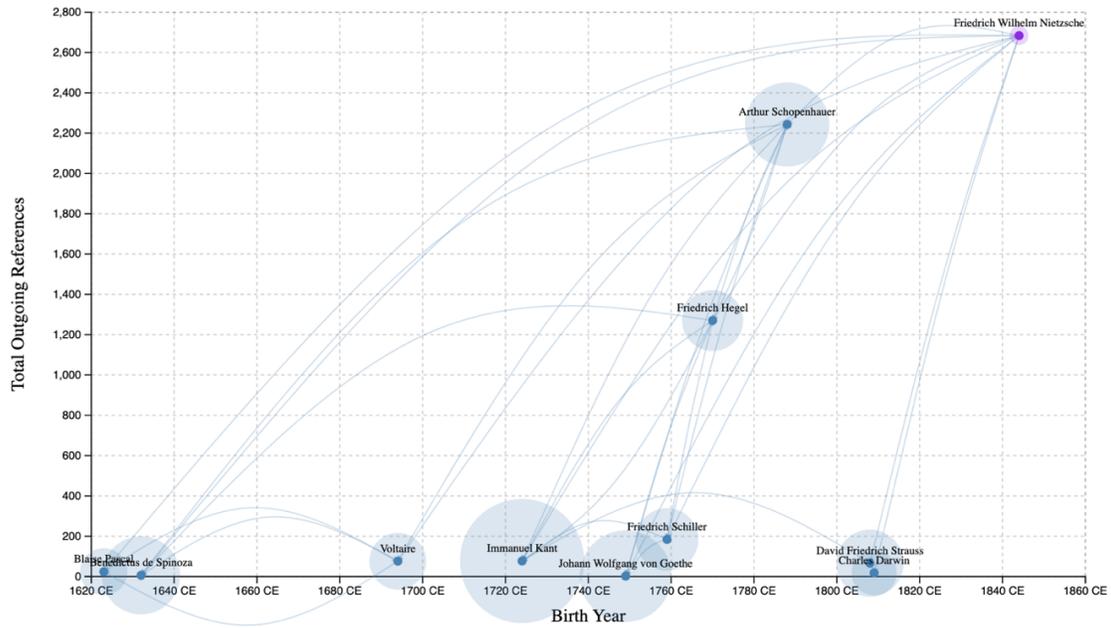

*Figure 29: Nietzsche's Position in the Reference Network*

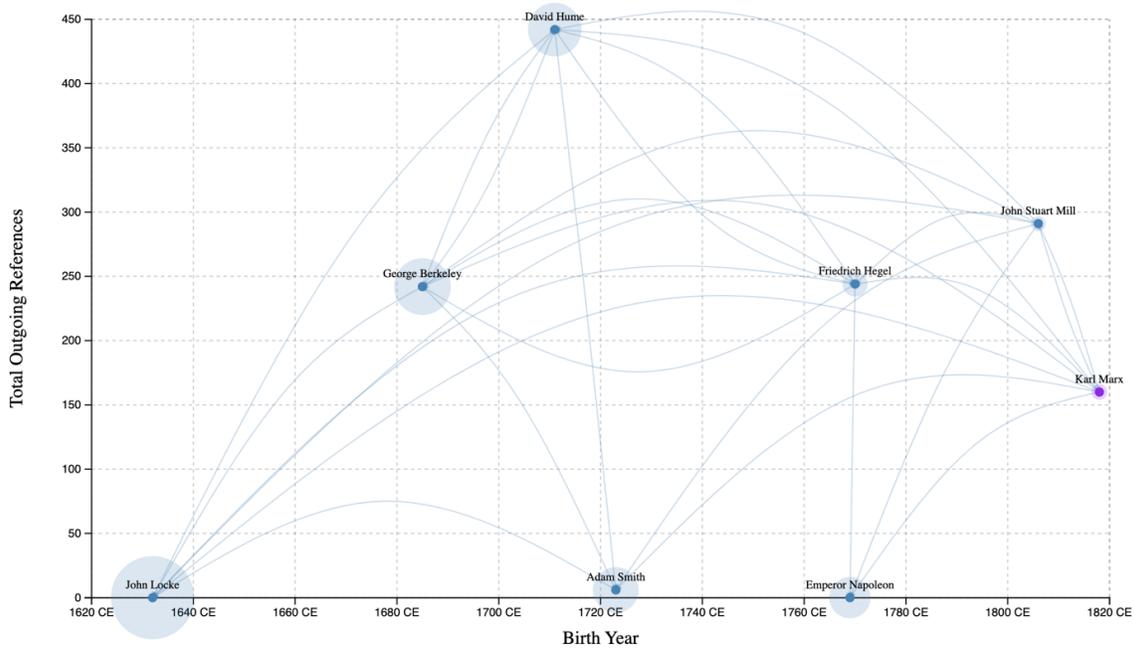

*Figure 30: Marx's Position in the Reference Network*

*5.5 The Giant Shoulders of Newton*

Among all the highest impact thinkers in our network, Newton is the least associated with the standard philosophical canon (see Fig. 31). This is somewhat curious—he worked on similar problems to Leibniz, and, up until his time, physics was quite connected with philosophy. Outside of our dataset, Newton also produced many purely philosophical works, which engaged with the rest of the philosophical canon in a more traditional manner. It is likely that the sheer practical utility and success of his theory of physics, along with his discovery of calculus, led us to retroactively categorize his contributions as an entirely distinct area from philosophy.

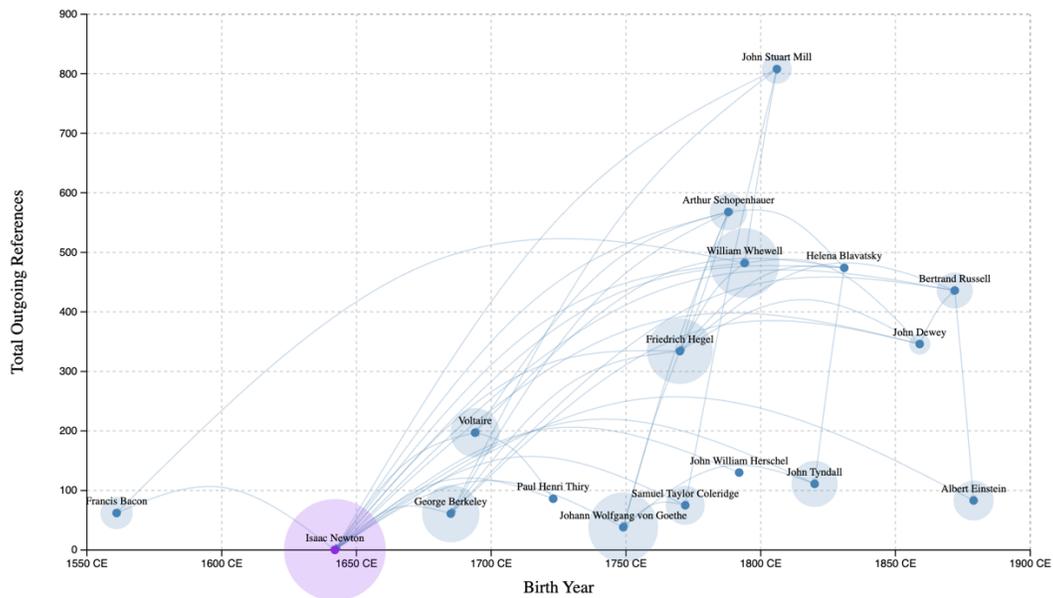

*Figure 31: Newton's Position in the Reference Network*

However, notions of Newtonian physics are still characterized as philosophical by many circles. He held in the Principa Mathematica that "Absolute space, in its own nature, without relation to anything external, remains always similar and immovable" (Newton 1999, Book I, Scholium to Definitions). His rival, Leibniz, considered "space to be something merely relative, as time is; that I hold it to be an order of coexistences, as time is an order of successions" (Leibniz 2000, Fourth Paper, §4). Modern philosophers tended to lean toward Leibniz's view, or build off of

Kant's efforts in synthesizing them. Physicists unanimously sided with Newton until Einstein's theory of General Relativity, which shifted a scientific paradigm back toward a more Leibnizian view (Friedman 2001) (see Fig. 32, Fig. 33).

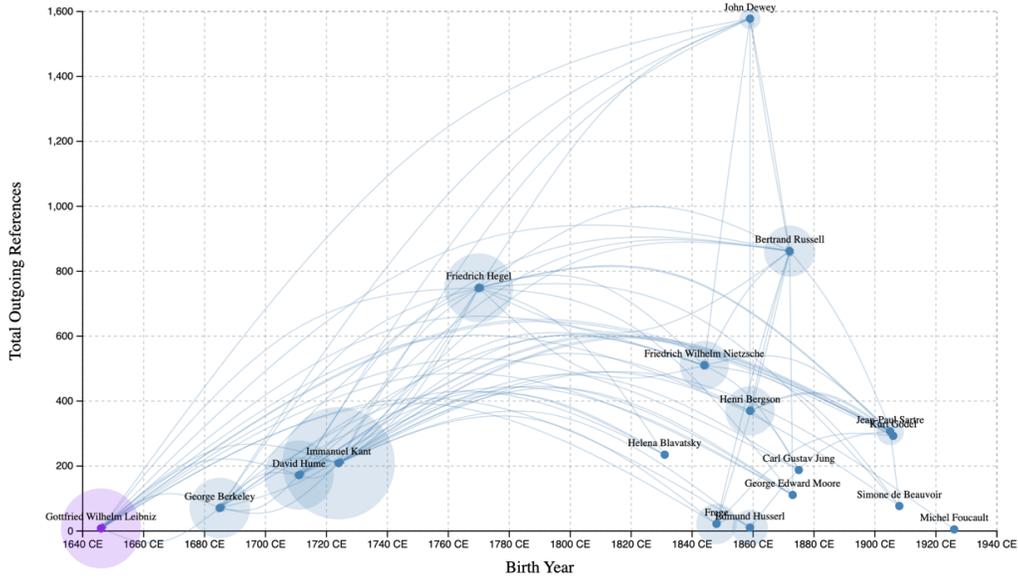

*Figure 32: Leibniz's Position in the Reference Network*

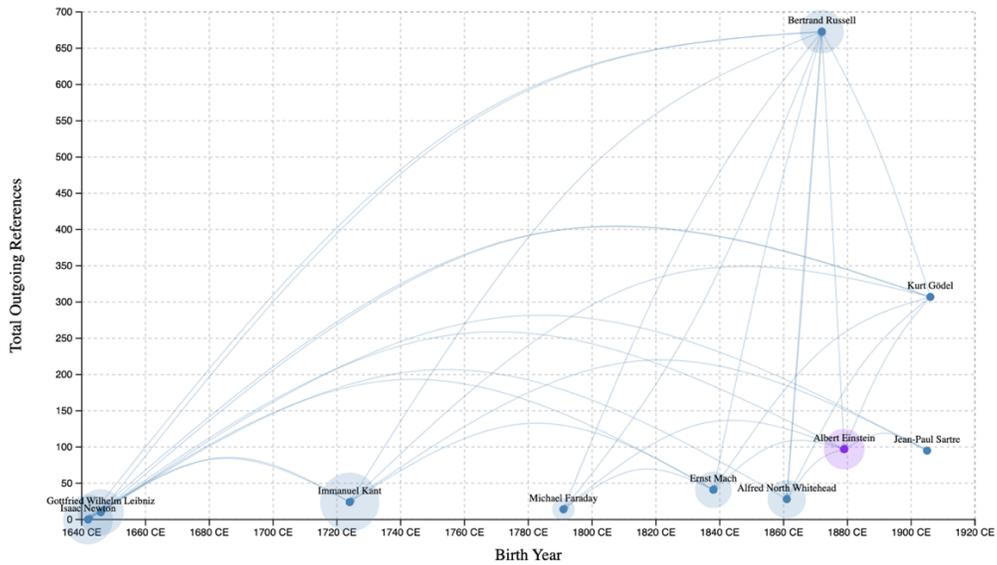

*Figure 33: Einstein's Position in the Reference Network*

This reflects an important distinction in the differences between scientific theories and philosophical: scientific theories are more widely adopted when they offer immediate, empirical, explanatory power, while philosophy can adopt views with more flexibility based on *a priori* justifications. The fifth postulate in Euclidean geometry had an analogous treatment: while in theoretical circles, mathematicians questioned its necessity, it provided a strong foundation for practical applications of geometry, especially within the Newtonian model of physics (see Fig. 34). Einstein's General Theory was the first major scientific work to take advantage of Non-Euclidean Geometry, which abandons the fifth postulate.

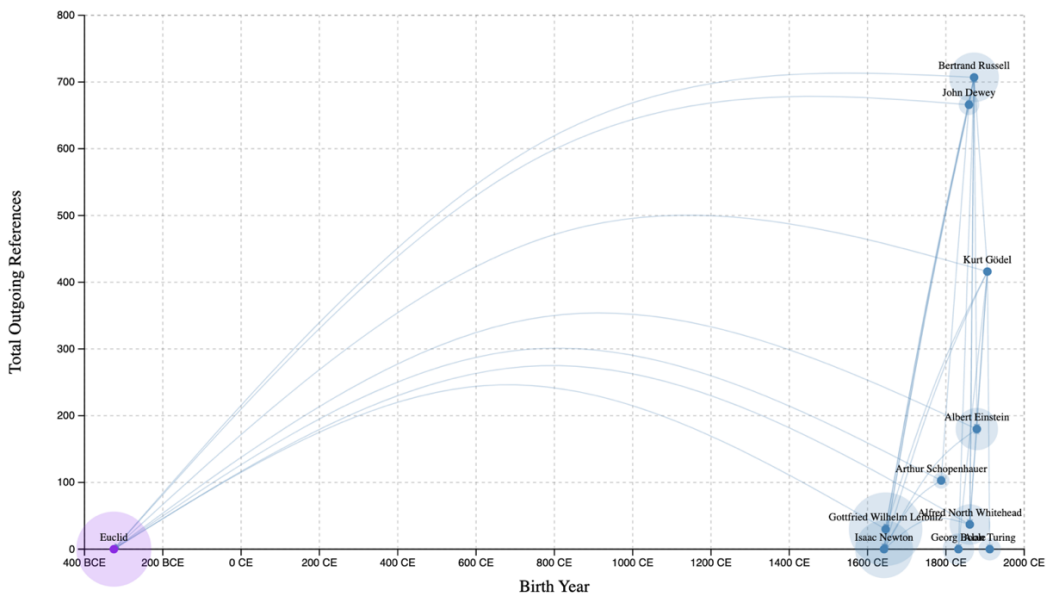

Figure 34: Euclid's Position in the Reference Network

5.6 The Descendants of Darwin

Darwin's position in our network is quantitatively similar to Newton, indicating their shared status as the most frequently referenced thinkers who are considered chiefly scientists, not philosophers (see Fig. 35). Like Newton, some of Darwin's lesser-known writings engage with important modern philosophers, such as Hume, as well as more literary figures, such as Goethe. Biology is arguably less connected to philosophy than physics, though Darwin's theories served as natural analogies in political, ethical, and religious debates, which frequently cited him. In Bertrand Russell's account, many of these extrapolations would not have pleased Darwin:

> "Darwin himself was a liberal, but his theories had consequences in some degree inimical to traditional liberalism… The conception of organism came to be thought the key to both scientific and philosophical explanations of natural laws, and the atomic thinking of the eighteenth century came to be regarded as out of date. This point of view has at last influenced even theoretical physics. In politics it leads naturally to emphasis upon the community as opposed to the individual… also with nationalism, which can appeal to the Darwinian doctrine of survival of the fittest applied, not to individuals, but to nations." (Russell 1945, 763).

Darwin's ideas have certainly survived. But, in the process of being passed down across generations, they have mutated, evolved, and adapted to different environments, becoming quite distinct from their shared ancestor.

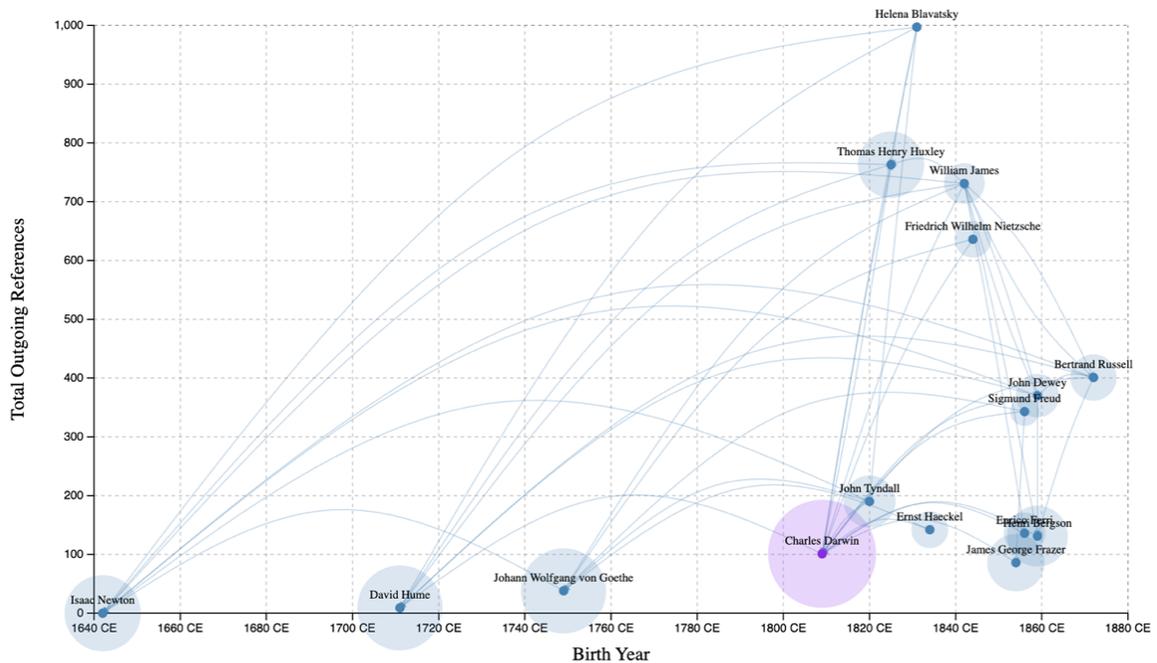

*Figure 35: Charles Darwin's Position in the Reference Network*

*5.7 The Seeds of New Disciplines*

Between the 1850s and 1920s, our network shows common patterns among thinkers associated with the development of distinct disciplines. The pioneering psychologists, Jung, Freud, and William James, all frequently referenced the modern and ancient philosophical canon (see Fig. 36). They are themselves chiefly referenced by each other, but also by many less-known psychologists who rarely reference the philosophical canon which gave rise to psychology. Likewise, Marx and Engels reference canonical works frequently, but receive the most references from essayists, activists, and predecessors who are more detached (see Fig. 37). We also see the same pattern with John Dewey, who receives the most incoming references from obscure educational theorists (see Fig. 38); the logicians, such as Wittgenstein, Whitehead, and Frege (see Fig. 39); mathematicians, such as Kurt Gödel, Alan Turing, and Georg Boole (see Fig. 30); and with Russian novelists, such as Tolstoy and Dostoevsky (see Fig. 40). For each of these communities, the major figures reference each other, while the minor figures appear isolated from the larger thread. It is fitting that Bertrand Russell, who played a crucial role in informing our own analysis, acts as a major bridge in this portion of the network, with a high number of outgoing and incoming references to many major philosophers in this area of the network (see Fig. 41).

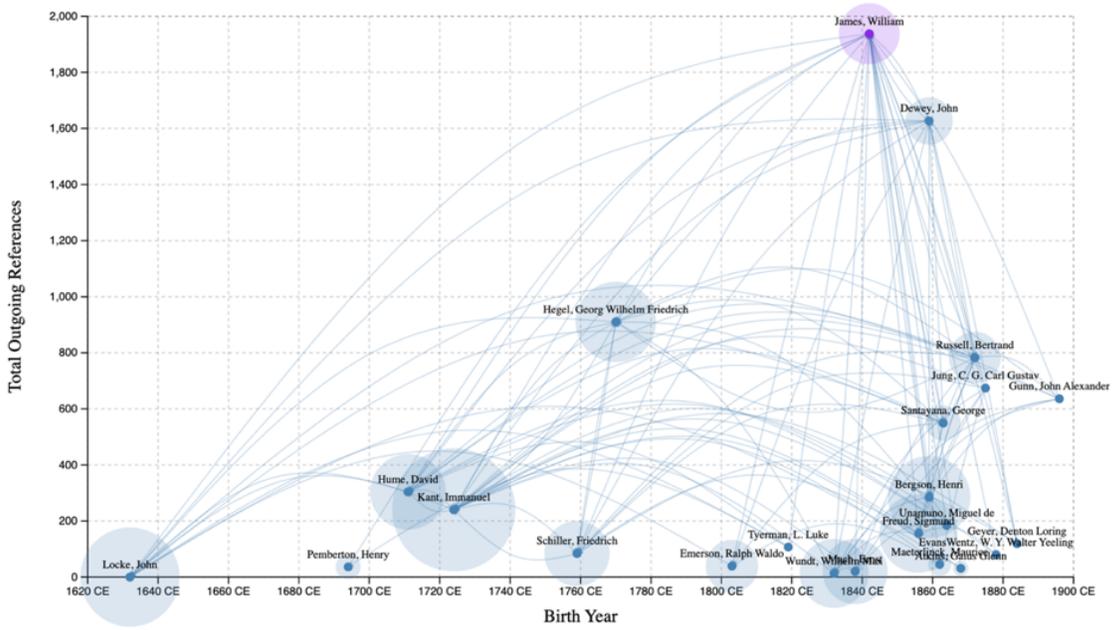

*Figure 36: William James' Position in the Reference Network*

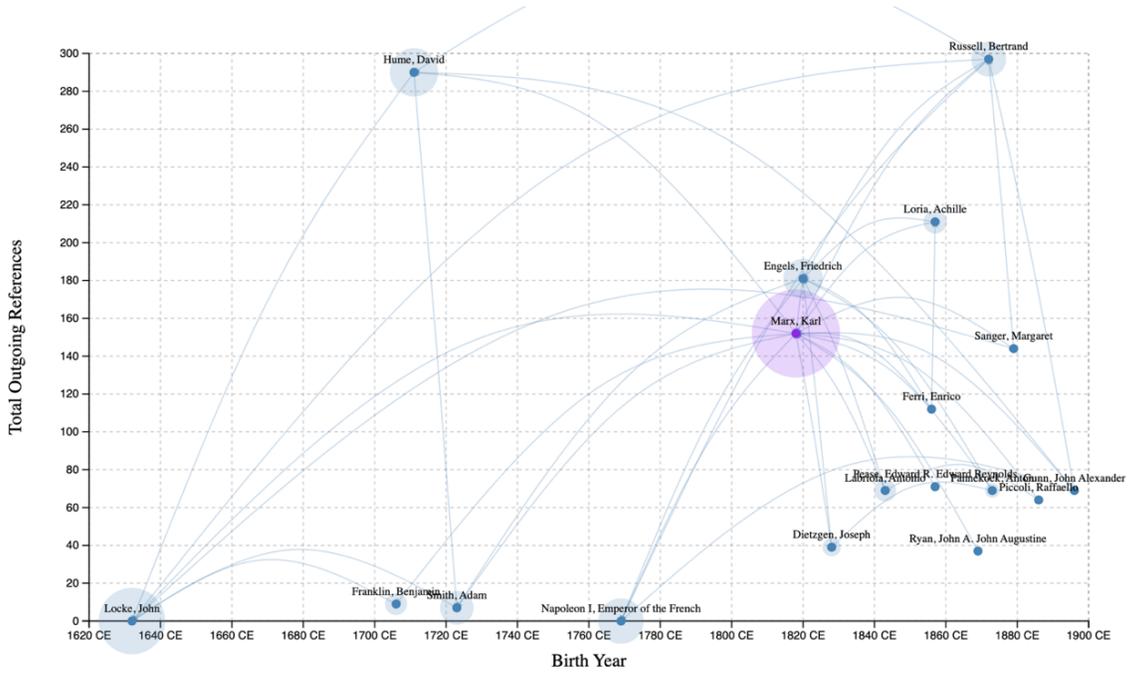

*Figure 37: Karl Marx's Position in the Reference Network*

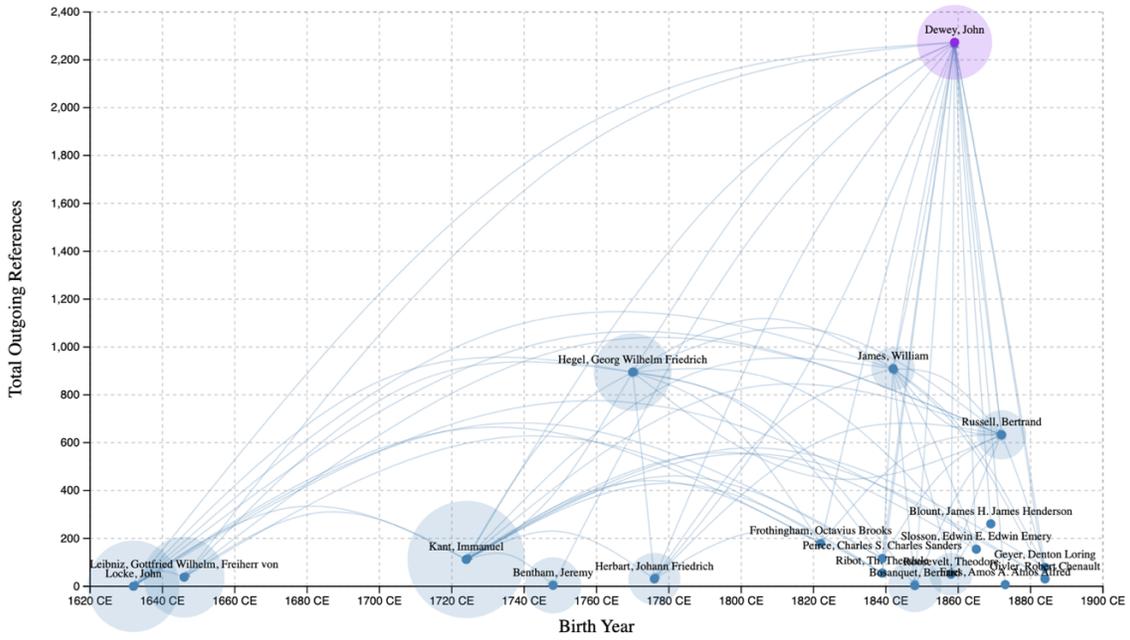

Figure 38: John Dewey's Position in the Reference Network

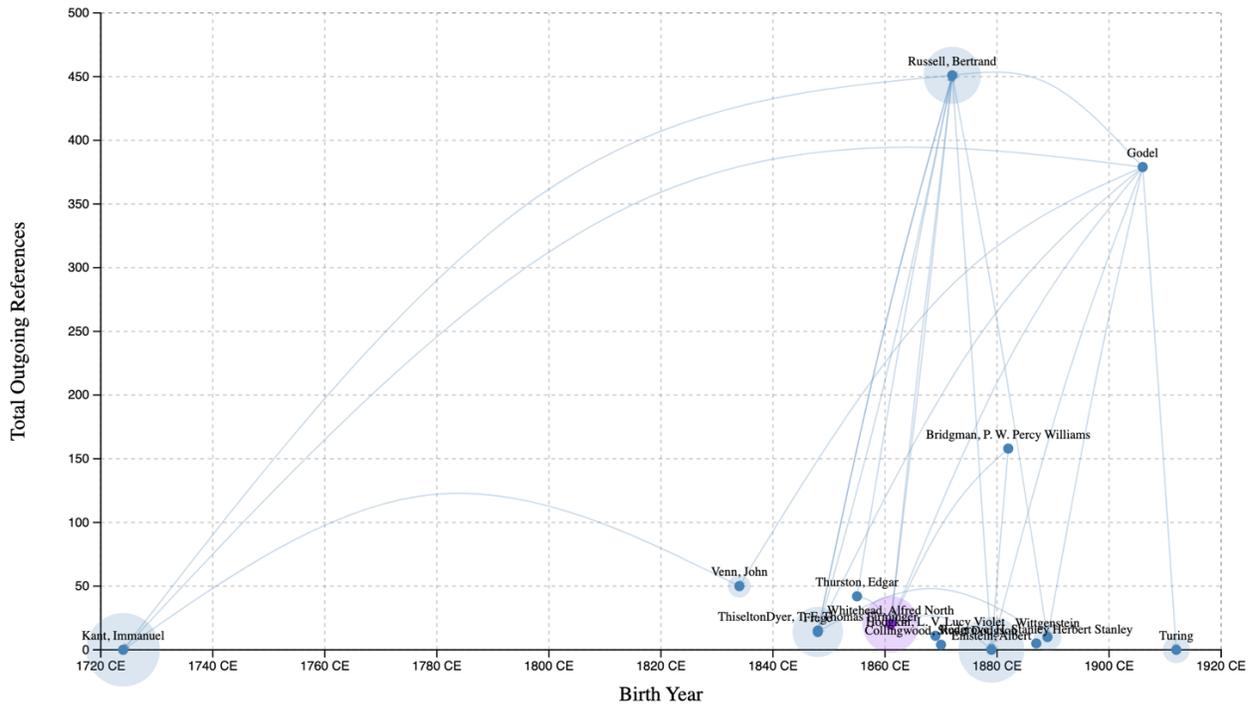

Figure 39: Mathematicians and Logicians in the Reference Network

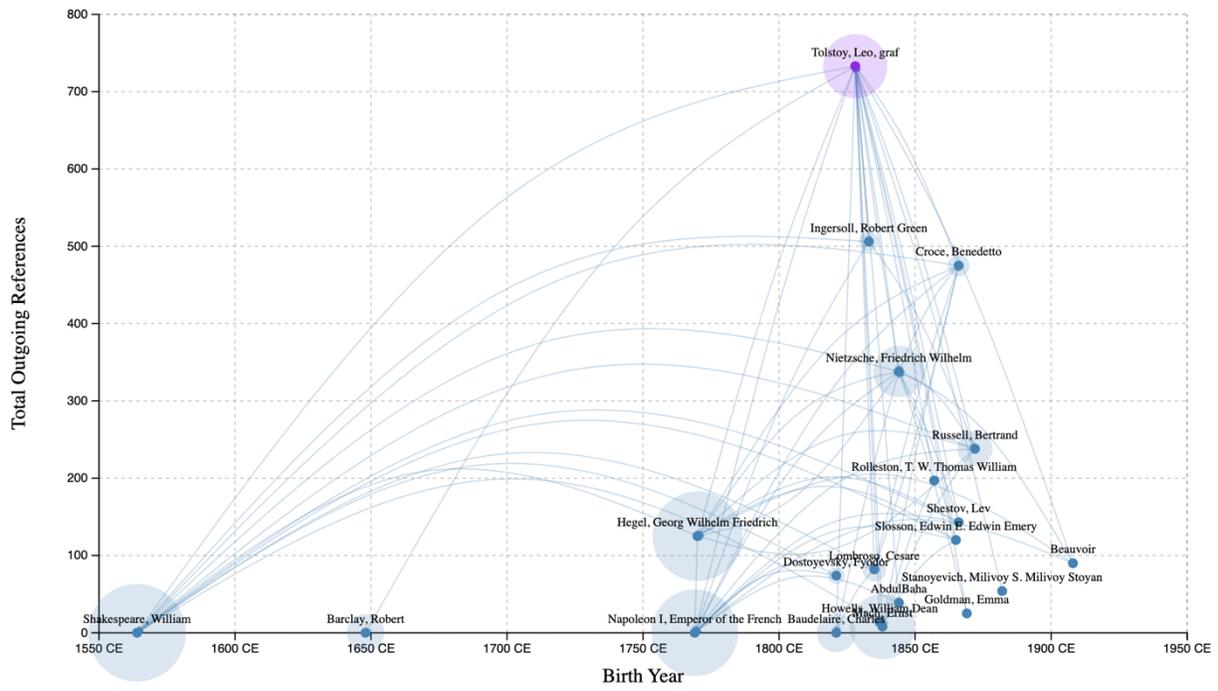

*Figure 40: Tolstoy in the Reference Network*

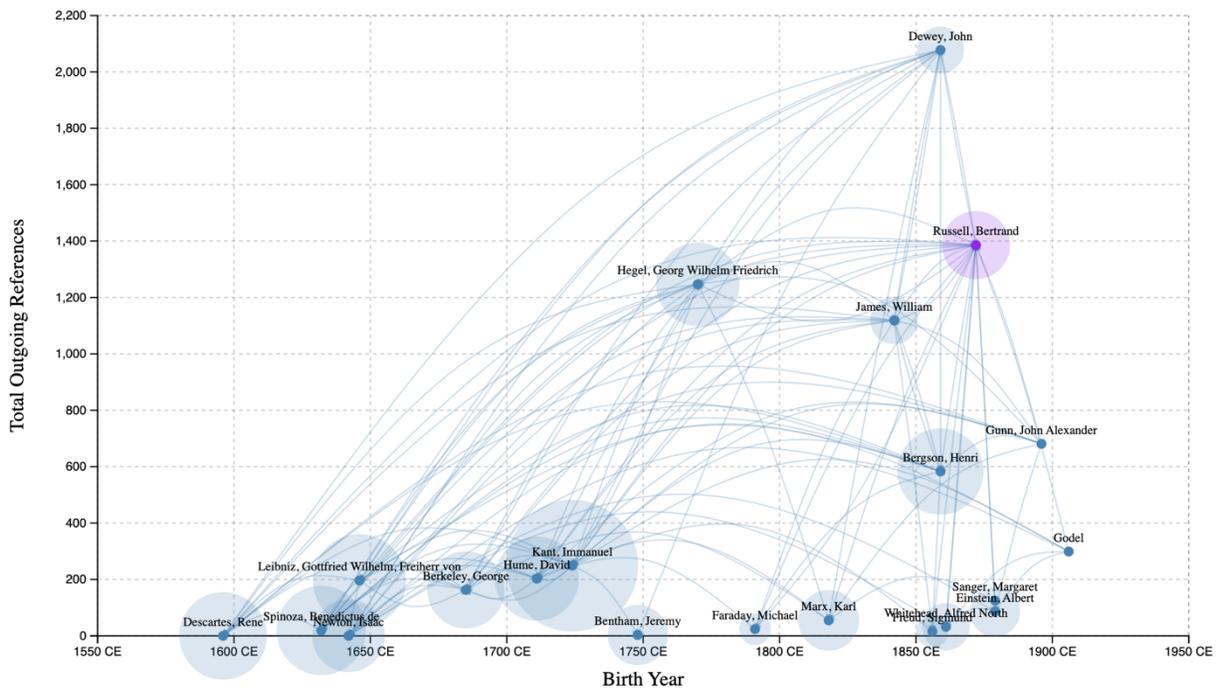

*Figure 41: Bertrand Russell in the Reference Network*

**Conclusion & Future Work**

Our research presents the largest computational analysis of philosophical reference networks to date—covering 2,245 texts and 294,970 references from over 1,000 authors. Our findings validate many long-held understandings in philosophical scholarship—such as the centrality of Plato, Aristotle, and Kant—while also offering new quantitative detail about their relative influence, both in terms of their overall impact as well as their impact within specific subdisciplines of philosophy.

Our reference network could be expanded to include any author with surviving works—including contemporary authors with non-academic writings, philosophers with works in the Internet Archive, or philosophers from entirely separate disciplines. While some additional methods would be needed, incorporating multiple languages within our reference network would provide interesting insights into how intellectual lineages across cultures have interacted with each other. In addition, by exclusively using incoming references, we could expand our reference network to include authors without surviving works—such as pre-Socratics, religious figures, or even fictional and mythological characters. There are also many influential individuals in our dataset—such as Virgil, Shakespeare, or Plotinus—whom we did not have the time to analyze in-depth, but have large scholarly communities dedicated to interpreting their works. Scholars with greater familiarity with their works could provide a much in-depth analysis of their position within our reference network, either through directly accessing our data or by using our visualization tool.

In addition to topic classification, the semantic context of each reference can be leveraged through methods such as stylometric analysis (Jockers, 2013) and sentiment analysis (Reagan et al, 2016) to more precisely measure the nature of references within our historical network. For instance, we could find how often philosophers agree on various topics or detect how often philosophers mention certain arguments. Large Language Models could serve as effective tools to characterize references without the need for a specialized model for each classification task. LLMs could also allow us to conduct a more thorough context disambiguation, either to better eliminate false matches or to identify all instances of in-text references made by an editor rather than the original author of a text. We also see a potential area of research in analyzing synthetic reference networks produced from LLMs. In preliminary work, we derived synthetic networks by measuring how often an LLM mentions an author when prompted with context of another author. These synthetic networks appear to represent the differences in how LLMs represent philosophical networks, which could serve researchers interested in evaluating their historical accuracy.

Beyond our analysis, we demonstrate that extracting in-text references—rather than formal citations—offers a promising methodology for extending approaches from citation analysis into new domains. Researchers can apply these methods to any collection of historical texts, offering a new way to understand the relationships between individuals and their historical network.

Works Cited

Alfano, M., 'A semantic-network approach to the history of philosophy, or, what does Nietzsche talk about when he talks about emotion?', Philosophy and Other Thoughts (2017)

Barmpounis, E.-K., Pavlopoulos, J., Louridas, P., and Konstantina, D., 'Recognition and analysis of the proceedings of the Greek Parliament after WWII', Digital Humanities Quarterly, 18/1 (2024)

Baschera, L. (2009) 'Chapter Six. Aristotle and Scholasticism', in Torrance Kirby, E. and Campi, E. (eds) *A Companion to Peter Martyr Vermigli*. Leiden: Brill (Brill's Companions to the Christian Tradition, 16), pp. 133–159. https://doi.org/10.1163/ej.9789004175549.i-542.34

Blei, D. M., Ng, A. Y., and Jordan, M. I., 'Latent Dirichlet allocation', Journal of Machine Learning Research, 3 (2003): 993-1022

Blondel, V.D., Guillaume, J.-L., Lambiotte, R. and Lefebvre, E. (2008) 'Fast unfolding of communities in large networks', *Journal of Statistical Mechanics: Theory and Experiment*, 2008(10), P10008. https://doi.org/10.1088/1742-5468/2008/10/P10008

Bommasani, Rishi, et al. "On the Opportunities and Risks of Foundation Models." *arXiv preprint arXiv:2108.07258* (2021).

Bostock, M., Ogievetsky, V. and Heer, J. (2011) 'D3: Data-driven documents', *IEEE Transactions on Visualization and Computer Graphics*, 17(12), pp. 2301–2309. https://doi.org/10.1109/TVCG.2011.185

Brown, D.G., Hutchinson, R. and Lamb, C.E. (2024) *A systematic mapping review of algorithms for the detection of rhymes, from early digital humanities projects to the rise of large language models*. Waterloo: University of Waterloo. Available at: http://hdl.handle.net/10012/20723

Cary, P. (2008) *Inner Grace: Augustine in the Traditions of Plato and Paul*. Oxford University Press.

Chandra, R., & Ranjan, M. (2022). Artificial intelligence for topic modelling in Hindu philosophy: Mapping themes between the Upanishads and the Bhagavad Gita. PLOS ONE, 17(9), e0273476. https://doi.org/10.1371/journal.pone.0273476

Chen, C.-M., Ho, S.-Y., & Chang, C. (2023). A hierarchical topic analysis tool to facilitate digital humanities research. Aslib Journal of Information Management, 75(1), 1-19. https://doi.org/10.1108/AJIM-11-2021-0325

Chi, PS., Conix, S. Measuring the isolation of research topics in philosophy. *Scientometrics* **127**, 1669–1696 (2022). https://doi.org/10.1007/s11192-022-04276-y.

Colish, M. L., St. Thomas Aquinas in Historical Perspective: The Modern Period (Cambridge: Cambridge University Press, 2009)

Crawford, Kate. *Atlas of AI: Power, Politics, and the Planetary Costs of Artificial Intelligence*. Yale University Press, 2021.

Crenshaw, K. (1991) 'Mapping the margins: Intersectionality, identity politics, and violence against women of color', *Stanford Law Review*, 43(6), pp. 1241–1299. https://doi.org/10.2307/1229039

Devlin, J., Chang, M.-W., Lee, K., and Toutanova, K., 'BERT: Pre-training of deep bidirectional transformers for language understanding', Proceedings of the 2019 Conference of the North American Chapter of the Association for Computational Linguistics: Human Language Technologies, Volume 1 (2019): 4171-4186

Fischer, S., Knappen, J., and Teich, E., 'Using topic modelling to explore authors' research fields in a corpus of historical scientific English', Proceedings of the Digital Humanities Conference 2018 (2018)

Forster, Michael, 'Hegel's Dialectical Method', in Frederick C. Beiser (ed.), The Cambridge Companion to Hegel (Cambridge: Cambridge University Press, 1993)

Friedman, M. 2001. *Dynamics of Reason*. Stanford: CSLI Publications.

Gilson, E. (1956) *The Christian Philosophy of St. Thomas Aquinas*. New York: Random House.

Greene, D., O'Sullivan, J., and O'Reilly, D., 'Topic modelling literary interviews from The Paris Review', Digital Scholarship in the Humanities, 39/1 (2024): 142-153

Gutendex (n.d.) *Project Gutenberg API*. Available at: https://gutendex.com

Healy, K., 'A Co-Citation Network for Philosophy', Blog post (2013), https://kieranhealy.org/blog/archives/2013/06/18/a-co-citation-network-for-philosophy/

Hagberg, A., Swart, P. and S Chult, D. (2008) 'Exploring network structure, dynamics, and function using NetworkX', in *Proceedings of the 7th Python in Science Conference (SciPy2008)*, Pasadena, CA, pp. 11–15.

Hendrix, S. (2015) *Martin Luther: Visionary Reformer*. Yale University Press.

Herbelot, A., von Redecker, E., and Müller, J., 'Distributional techniques for philosophical enquiry', in K. Zervanou and A. van den Bosch (eds.), Proceedings of the 6th Workshop on Language Technology for Cultural Heritage, Social Sciences, and Humanities (Association for Computational Linguistics, 2012): 45-54

Hinrichs, U., Forlini, S., and Moynihan, B., 'In defense of sandcastles: Research thinking through visualization in digital humanities', Digital Scholarship in the Humanities, 34/S1 (2019): i80-i99

Hopkins, B., The Philosophy of Husserl (London: Routledge, 2011)

Israel, J., 'Enlightenment! Which Enlightenment?', Journal of the History of Ideas, 67/3 (2006): 523-545

Jelodar, H., et al., 'Latent Dirichlet allocation (LDA) and topic modeling: models, applications, a survey', Multimedia Tools and Applications, 78 (2019): 15169-15211

Jockers, Matthew L. *Macroanalysis: Digital Methods and Literary History*. University of Illinois Press, 2013.

Kamtekar, R., 'Levels of Argument: A Comparative Study of Plato's Republic and Aristotle's Nicomachean Ethics', Ancient Philosophy, 36/1 (2016): 214-221

Kant, I. (1781/1998) *Critique of Pure Reason*, trans. P. Guyer and A.W. Wood. Cambridge: Cambridge University Press.

Köntges, Thomas. (2020). Measuring Philosophy in the First Thousand Years of Greek Literature. Digital Classics Online, 6(2). https://doi.org/10.11588/dco.2020.2.73197

Koroteev, M. V., 'BERT: A Review of Applications in Natural Language Processing and Understanding', arXiv:2103.11943 [cs.CL] (2021)

Kennedy, J. (2020) 'Kurt Gödel', in Zalta, E.N. (ed.) *The Stanford Encyclopedia of Philosophy* (Winter 2020 Edition). Available at: https://plato.stanford.edu/archives/win2020/entries/goedel/ (Accessed: 1 April 2025).

Kwon, H.-C., and Shim, K.-H., 'An improved method of character network analysis for literary criticism: A case study of Hamlet', International Journal of Contents, 13/3 (2017)

Lee, D. D., and Seung, H. S., 'Learning the parts of objects by non-negative matrix factorization', Nature, 401/6755 (1999): 788-791

Lee, J. S. Y., and Webster, C., 'Conversational networks: Prophets and kings in the Old Testament', Digital Scholarship in the Humanities, 39/4 (2024): 1043-1063

Leibniz, G.W. 2000. *The Leibniz-Clarke Correspondence*. Edited by R. Ariew. Indianapolis: Hackett Publishing.

Leiter, B., 'Nietzsche Against the Philosophical Canon', University of Chicago Public Law & Legal Theory Working Paper No. 438 (2013)

Lucia Giagnolini, Marilena Daquino, Francesca Mambelli, Francesca Tomasi, Exploratory methods for relation discovery in archival data, *Digital Scholarship in the Humanities*, Volume 38, Issue 1, April 2023, Pages 111–126, https://doi.org/10.1093/llc/fqac036

Petrovich, E., Viola, M. Mapping the philosophy and neuroscience nexus through citation analysis. *Euro Jnl Phil Sci* **14**, 60 (2024). https://doi.org/10.1007/s13194-024-00621-5.

Nietzsche, F. (1874) *Schopenhauer as Educator*. In: *Untimely Meditations*, Part III. Trans. by Gersimon and R.J. Hollingdale. Available at: https://la.utexas.edu/users/hcleaver/330T/350kPEENietzscheSchopenTable.pdf (Accessed: 1 April 2025).

Newton, I. 1999. *The Principia: Mathematical Principles of Natural Philosophy*. Translated by I. B. Cohen and A. Whitman. Berkeley: University of California Press.

Malaterre, C., Lareau, F. The early days of contemporary philosophy of science: novel insights from machine translation and topic-modeling of non-parallel multilingual corpora. *Synthese* **200**, 242 (2022). https://doi.org/10.1007/s11229-022-03722-x.

MacDonald Ross, G. (ed.), Kant and His Influence (London: Thoemmes Continuum, 2006)

Morris, K., and Preti, C. (eds.), Early analytic philosophy: An inclusive reader with commentary (London: Bloomsbury Academic, 2023)

Pachayappan, M., and Venkatesakumar, R., 'A graph theory-based systematic literature network analysis', Theoretical Economics Letters, 8/5 (2018)

Reagan, Andrew J., et al. "The emotional arcs of stories are dominated by six basic shapes." *EPJ Data Science* 5.1 (2016): 1–10.

Ruegg, C., and Lee, J. J., 'Epic social networks and Eve's centrality in Milton's Paradise Lost', Digital Scholarship in the Humanities, 35/1 (2020): 146-159

Russell, B. (1945) *A History of Western Philosophy*. New York: Simon & Schuster.

Satlow, M. L., and Sperling, M., 'Social network analysis of the Babylonian Talmud', Digital Scholarship in the Humanities (2024)

S Sanh, V., Debut, L., Chaumond, J. and Wolf, T. (2019) 'DistilBERT, a distilled version of BERT: smaller, faster, cheaper and lighter', *arXiv preprint*. Available at: https://arxiv.org/abs/1910.01108 (Accessed: 1 April 2025).

Sarkar, S., Feng, D., & Karmaker Santu, S. K. (2023). Zero-Shot Multi-Label Topic Inference with Sentence Encoders and LLMs. In H. Bouamor, J. Pino, & K. Bali (Eds.), Proceedings of the 2023 Conference on Empirical Methods in Natural Language Processing (pp. 16218-16233). Association for Computational Linguistics. https://doi.org/10.18653/v1/2023.emnlp-main.1007.

Szaszi, B., Habibnia, H., Tan, J., Hauser, O.P. and Jachimowicz, J.M. (2024) 'Selective insensitivity to income held by the richest', *PNAS Nexus*, 3, pgae333. https://doi.org/10.1093/pnasnexus/pgae333

Schöch, Christof. (2017). Topic Modeling Genre: An Exploration of French Classical and Enlightenment Drama. Digital Humanities Quarterly, 11(2). http://www.digitalhumanities.org/dhq/vol/11/2/000291/000291.html

Shapiro, L., 'Revisiting the early modern philosophical canon', Journal of the American Philosophical Association (2016)

Treadgold, W. (1997) *A History of the Byzantine State and Society*. Stanford, CA: Stanford University Press.

United States. (1776) *Declaration of Independence*. National Archives. Available at: https://www.archives.gov/founding-docs/declaration-transcript (Accessed: 1 April 2025).

Varga, R., and Bornhofen, S., 'Graph-based modelling of prosopographical datasets: Case study: Romans 1by1', Digital Humanities Quarterly, 18/2 (2024)

van Boven, G., and Bloem, J., 'Domain-specific Evaluation of Word Embeddings for Philosophical Text using Direct Intrinsic Evaluation', Proceedings of the 2nd International Workshop on Natural Language Processing for Digital Humanities (2022)

Wang, C., Castellón, I., and Comelles, E., 'Linguistic analysis of datasets for semantic textual similarity', Digital Scholarship in the Humanities, 35/2 (2020): 471-484

Wasserman, S., and Faust, K., Social Network Analysis: Methods and Applications (Cambridge: Cambridge University Press, 1994)

Weatherson, B., A history of philosophy journals: Volume 1: Evidence from topic modeling, 1876-2013 (Ann Arbor: Maize Books, 2020)

Whitehead, A.N. (1929) *Process and Reality: An Essay in Cosmology*. New York: Macmillan.

Wilkens, M., Mimno, D. and Walsh, M. (2023) *BERT for Humanists*. Cornell University. Available at: https://www.bertforhumanists.org (Accessed: 1 April 2025).

Zalta, E.N. (ed.) (n.d.) *The Stanford Encyclopedia of Philosophy*. Stanford University. Available at: https://plato.stanford.edu (Accessed: 1 April 2025).

Appendix

Scripts for our data processing, reference extraction, topic modeling, and network analysis are available on our GitHub repository: https://github.com/ogreowl/nlp-phil

Our visualization tool, PhilBERT, is publicly accessible at: https://ogreowl.github.io/PhilBERT/